\documentclass[10pt,a4paper]{article}
\usepackage[top=1.8cm, bottom=1.8cm, left=1.5cm, right=1.5cm]{geometry}
\usepackage[utf8]{inputenc}
\usepackage{amsmath}
\usepackage{amsfonts}
\usepackage{amssymb}
\usepackage{caption}
\usepackage{subcaption}
\usepackage{float}
\usepackage{graphicx}
\usepackage{epstopdf}
\usepackage{hyperref}
\usepackage{multicol} 
\usepackage{authblk}

\title{
\begin{center}
{Axially Symmetric Accretion of Fractal Medium onto Rotating Black Holes and the emergence of the Acoustic Manifold}
\end{center}
}

\author[]{
Supriyo Majumder$^{1}$\thanks{email: supriyom@csr.res.in; Present address: UGC DAE Consortium for Scitific Research, Indore 452001, India.}, 
\ 
Tapas K. Das$^{2}$\thanks{email: tapas@hri.res.in} 
\ and 
Sankhasubhra Nag$^{3}$\thanks{email: sankha@sncwgs.ac.in}
\\
$^{1}$Barasat Government College, Kolkata 700124, India.\\
$^{2}$Harish-Chandra Research Institute, Allahabad 211019, India.\\
$^{3}$Sarojini Naidu College for Women, Kolkata 700028, India.\\
}

\begin{document}

\maketitle

\begin{abstract}
For three different geometric configurations and two different thermodynamic equations of state, low angular momentum, multi-transonic, axially symmetric accretion flow of matter having fractional dimension of mass distribution onto a rotating black hole has been studied by employing certain post-Newtonian pseudo-Kerr black hole potential. Such task has been accomplished mathematically by mapping the fractal nature of accreted medium onto its continuum counterpart. The difference between spin dependence of accretion dynamics of the fractal medium and the continuous medium has been highlighted. By employing a time dependent linear perturbation scheme, it has been demonstrated that accretion of matter with fractional dimension of density distribution can be considered as a natural example of classical analogue model. The corresponding acoustic surface gravity has been estimated in terms of accretion variables. The value of the surface gravity changes as the accreted matter makes a transition from its fractal nature to the corresponding continuum distribution.
\end{abstract}


\begin{multicols}{2}

\section{INTRODUCTION}
\paragraph*{}
The multi transonic behaviour, and the formation of the standing shock as a consecuence of such profile, has been studied by several authors for black hole accretion under the influence of various post-Newtonian black hole potentials (see Liang \& Thomson 1980\cite{Liang & Thomson 1980}; Abramowicz \& Zurek 1981\cite{Abramowicz et al. 1981}; Muchotrzeb \& Paczynski 1982\cite{Muchotrzeb & Paczynski 1982}; Muchotrzeb 1983\cite{Muchotrzeb 1983}; Muchotrzeb-Czerny 1986\cite{Muchotrzeb 1986}; Blaes 1987\cite{Blaes 1987}; Abramowicz \& Kato 1989\cite{Abramowicz et al. 1989}; Chakrabarti 1989\cite{Chakrabarti 1989}; Das 2002\cite{Das 2002}; Das Pendharkar \& Mitra 2003\cite{Das et al. 2003}; Nag et al. 2012\cite{Nag et al. 2012}; Saha et al. 2016\cite{Saha et al. 2016}). The connection between the mathematical equations governing such flow and the set of first order differential equations describing autonomous dynamical systems has recently been established by several works (Ray \& Bhattacharjee 2002\cite{Ray et al. 2002}; Ray 2003a\cite{Ray 2003a}, b\cite{Ray 2003b}; Ray \& Bhattacharjee 2005b\cite{Ray et al. 2005b}, a\cite{Ray et al. 2005a}; Ray \& Bhattacharjee 2006\cite{Ray et al. 2006}, 2007a\cite{Ray et al. 2007}; Bhattacharjee \& Ray 2007\cite{Bhattacharjee et al. 2007}; Ray \& Bhattacharjee 2007b\cite{Ray et al. 2007b}; Chaudhury Ray \& Das\cite{Chaudhury et al. 2006}; Nag et al. 2012\cite{Nag et al. 2012}; Saha et al. 2016\cite{Saha et al. 2016}). 

\paragraph*{}
Quite recently, it has been argued that accreting black holes may be considered as classical analogue gravity models, and for axially symmetric accretion under the framework of pseudo-Newtonian black hole potentials, a linear perturbation scheme may be developed to study the emergence of curved acoustic gravity embedded within such flow structure. The corresponding acoustic surface gravity has computed leading to the understanding of the analogue Hawking (see Novello et. al. 2002d\cite{Novello2002} and references therein) like effects (Nag et al. 2012\cite{Nag et al. 2012}; Bili\'{c} et al. 2014\cite{Bilic et al. 2014}; Saha et al. 2016\cite{Saha et al. 2016}). Recent systematic studies of low angular momentum inviscid black hole accretion thus leads to the understanding of various astrophysical phenomena as well as to the proper realization of analogue gravity effects as observed within the non-quantum fluids.

\paragraph*{}
The aforementioned works, quite naturally, assumes the accreting fluid to be a continuum. There are, however, recent observational indications that, interstellar matter (ISM) may have certain ``clumpish fractal" structure (Langer et al. 1995\cite{Langer et al. 1995}; Crovisier, Dickey \& Kaz\`{e}s 1985\cite{Crovisier et al. 1985}; Faison et al. 1998\cite{Faison et al. 1998}, Hill et al. 2005\cite{Hill et al. 2005}). These observational findings, as well as some related proposals (Falgarone et al. 1991\cite{Falgarone et al. 1991}, Larson 1981\cite{Larson 1981}, Falgarone et al. 1992\cite{Falgarone et al. 1992}, Zimmermann \& Stutzki 1992\cite{Zimmermann & Stutzki 1992}, Heithausen et al. 1998\cite{Heithausen et al. 1998}), tempted a group of workers to introduce a model for transonic accretion in fractal media (Roy 2007\cite{Roy 2007}; Roy \& Ray 2007\cite{Roy et al. 2007}; Roy \& Ray 2009\cite{Roy et al. 2009}). By mapping the dynamics of fractal media onto the continuum space (Tarasov 2005c\cite{Tarasov 2005c}, M. Ostoja-Starzewski et al. 2013\cite{M. Ostoja-Starzewski et al. 2013}), Roy \& Ray 2009\cite{Roy et al. 2009}, presented a detailed analysis of the multi-transonic flow profile for accretion onto a non-rotating black hole using Paczy\'{n}ski \& Wiita 1980\cite{Paczynski and Wiita 1980} pseudo-Newtonian black hole potential considering the case that accretion is balanced vertically under hydrostatic equilibrium (i.e. Vertical Equilibrium disc height model). The fractal medium was approximated by continuum and the mass continuity equation was suitably tinkered by introducing some numerical factor in the power of the length scale to mimic the fractal medium. The modification factor is related with the fractional integration to be carried out on the continuum to match with the result obtained by performing the integration in fractal medium (within  some numerical factor) as it was already prescribed in the literature (Ren et al. 2003\cite{Ren et al. 2003}; Tarasov 2004\cite{Tarasov 2004}). 

\paragraph*{}
It is however, widely believed that most of the astrophysical black holes are of Kerr type (Brenneman 2013\cite{Brenneman 2013}; Buliga et al. 2011\cite{Buliga et al. 2011}; Daly 2011\cite{Daly 2011}; Dauser et al. 2010\cite{Dauser et al. 2010}; Dotti et al. 2013\cite{Dotti et al. 2013}; Fabian et al. 2014\cite{Fabian et al. 2014}; Garofalo 2013\cite{Garofalo 2013}; Healy et al. 2014\cite{Healy et al. 2014}; Jiang et al. 2014\cite{Jiang et al. 2014}; Kato et al. 2010\cite{Kato et al. 2010}; Mart\`{i}nez-Sansigre and Rawlings, 2011\cite{Martinez-Sansigre and Rawlings 2011}; McClintock et al. 2011\cite{McClintock et al. 2011}; McKinney et al. 2013\cite{McKinney et al. 2013}; Miller et al. 2009\cite{Miller et al. 2009}; Nemmen and Tchekhovskoy 2014\cite{Nemmen and Tchekhovskoy 2014}; Nixon et al. 2011\cite{Nixon et al. 2011}; Reynolds et al. 2012 \cite{Reynolds et al. 2012}; Sesana et al. 2014\cite{Sesana et al. 2014}; Tchekhovskoy and McKinney 2012\cite{Tchekhovskoy and McKinney 2012}; Tchekhovskoy et al. 2010\cite{Tchekhovskoy et al. 2010}; Ziolkowski 2010\cite{Ziolkowski 2010}; Saha et al. 2016\cite{Saha et al. 2016}).  On the other hand, already in the literature there is a host of prescriptions (Artemova et al. 1996\cite{Artemova et al. 1996}; Chakrabarti and Khanna 1992\cite{Chakrabarti and Khanna 1992}; Chakrabarti and Mondal 2006\cite{Chakrabarti and Mondal 2006}; Ghosh and Mukhopadhyay 2007\cite{Ghosh and Mukhopadhyay 2007}; Ghosh et al. 2014 \cite{Ghosh et al. 2014}; Karas and Abramowicz 2014\cite{Karas and Abramowicz 2014}; Lovas 1998\cite{Lovas 1998}; Mukhopadhyay 2002\cite{Mukhopadhyay 2002}; Semer\~{A}ak and Karas 1999\cite{SemerAak and Karas 1999}) regarding suitable pseudo-Kerr potentials within the post-Newtonian framework.

\paragraph*{}
In this present work, we thus intends to study how the black hole spin influences the dynamics of the axially symmetric accretion of the fractal medium. To accomplish such task, we would like to study the accretion phenomena using a post-Newtonian pseudo-Kerr black hole potential. A number of such potentials exist in the literature (\cite{Artemova et al. 1996, Chakrabarti and Khanna 1992, Chakrabarti and Mondal 2006, Ghosh and Mukhopadhyay 2007, Ghosh et al. 2014, Karas and Abramowicz 2014, Lovas 1998, Mukhopadhyay 2002, SemerAak and Karas 1999}), among which, we pick up the potential proposed by Artemova et al. 1996.

\paragraph*{}
The expression for the free fall acceleration as provided by Artemova et al.(1996)\cite{Artemova et al. 1996} is
\begin{equation}\label{eq.65}
f=-\frac{1}{r^{2-\xi} (r-r_{1})^\xi} .
\end{equation}
Here, $ r_{1} $ is the position of the event horizon. The length of the radial coordinate, $ r $,(measured along the equatorial plane of the flow), has been scaled in units of Schwarzschild radius, defined as $ r_{g} = {G M_{BH}}/{c^2} $ (with $ M_{BH} $ being the mass of the black hole, $ G $ the universal gravitational constant and $ c $ the velocity of light in vacuum). We use the system of unit where $c=G=M_{BH}=1$. The potential corresponding to the above acceleration diverges at the event horizon. This position is determined by exact expression from general relativity (see Novikov \& Frolov 1989\cite{Novikov & Frolov 1989}).\\
In the above expression 
\begin{equation}\label{eq.66}
r_{1} = 1+(1-a^2)^{1/2} .
\end{equation}
\begin{equation}\label{eq.67}
r_{in} = 3+Z_{2}-[(3-Z_{1})(3+Z_{1}+2Z_{2})]^{1/2} ,
\end{equation}
\begin{equation}\label{eq.68}
Z_{1}=1+(1-a^2)^{1/3}[(1+a)^{1/3}+(1-a)^{1/3}] ,
\end{equation}
\begin{equation}\label{eq.69}
Z_{2}=(3a^2+Z_{1}^2)^{1/2} ,
\end{equation}
\begin{equation}\label{eq.70}
\xi = \frac{r_{in}}{r_{1}} -1 ;
\end{equation} where \textquoteleft{$ a $}\textquoteright \ is the spin parameter of the rotating black hole and $ a \in [-1,1) $ .

Explicit expression for the associated black hole potential ($ \phi_{ABN} = \phi $) is given as (Ying Wang and Xin Wu\cite{W.Ying et al. 2012}; Saha et al. 2016\cite{Saha et al. 2016}) :\\
\begin{equation}\label{eq.71}
\phi = -\frac{G M}{(\xi-1)r_{1}} \left[ \left( \frac{r}{r-r_{1}} \right)^{\xi-1} -1 \right] . 
\end{equation}

\paragraph*{}
The aforementioned potential has the simplest form (among all the proposed pseudo-Kerr black hole potentials) to deal with, yet it nicely mimics the astrophysics in the Kerr metric within a reasonable Newtonian setup, especially in simulating the multitransonic accretion flow around rotating black holes. 

\paragraph*{}
In this present work we would like to provide a fairly comprehensive treatment of the accretion flow of fractal medium onto a Kerr black hole. Not only we have studied the multi-transonic flow properties, we also study such accreting black holes from dynamical system point of view. Apart from that, by linear perturbing such flows, we examine the nature of the emergent acoustic geometry, thereby calculate the value of the acoustic surface gravity $ \kappa $ and demonstrate how $ \kappa $ gets influenced by the fractal nature of the medium, as well as by the black hole spin. We thus address the problem from astrophysical point of view as well as from perspective of the dynamical systems study and the analogue gravity phenomena. Such exhaustive treatment of a potentially new field i.e., a spinning black hole accreting fractal matter - has not been presented in literature yet. 
\\

The plan of the paper is as follows: 

In the text section, we will formulate and solve the continuity and the Euler equation for the flow of fractal medium by using a mapping of the fractal matter distribution onto its continuum counterpart, and will obtain the corresponding first integrals of motion, both for accretion governed by the polytropic as well as the isothermal equation of state. We then introduce a eigenvalue based linear perturbation scheme to map the stationary solutions of the aforementioned equations onto a set of first order differential equations for autonomous dynamical systems, to understand what would be the nature of the critical points encountered by the phase orbits corresponding to the flow. In subsequent sections, we pointed a full numerical study; the corresponding phase portrait of the multi-transonic accretion flow. Finally we show how the curved manifold of the acoustic geometry will emerge from the accreting black hole system and how the corresponding acoustic surface gravity can be computed as functions of flow variables, space-time metric elements, as well as the fractional dimension of the accreting matter.

\section{BASIC FORMALISM}

\paragraph*{}
The accreting material is assumed to possess low angular momentum, inviscid, axisymmetric flow and is considered to have `clumps' in it's structure. These self similar clumps can be approximated mathematically by a suitable fractal model which can imitate the actual physical system within some finite range of length scales. Following the procedure adopted in the literature (see Roy \& Ray 2009\cite{Roy et al. 2009}), the governing hydrodynamic equations for such axisymmetric fractal accretion, can be formulated using a homogenization scheme called dimensional regularization (Tarasov 2005a\cite{Tarasov 2005a}, 2005b\cite{Tarasov 2005b}, 2005c\cite{Tarasov 2005c}, Tarasov 2010\cite{Tarasov 2010}, M. Ostoja-Starzewski et al. 2013\cite{M. Ostoja-Starzewski et al. 2013}). With this transformation the fractional integrals (Ren et al. 2003\cite{Ren et al. 2003}), over fractal network are mapped to equivalent continuous integrals in which this fractal nature is embedded through fractional dimension of mass. Similar kind of approach can also be used here. The fractional infinitesimal length element in such fractional continuous accretion disc will be of the form,
\begin{equation}\label{eq.1}
d \overline{r} = \left( \frac{r}{l_{c}} \right)^{\Delta-1} dr , 
\end{equation}
where $ l_{c} $ is the characteristic inner length scale of the fractal medium, below which there is no density fluctuation i.e. continuum. Consequently the accretion disc has fractional dimension $ 2 \Delta $. Hence, a thin disc with volume density $ \rho $, radius, $ r $, and thickness, $ H $ will have mass,
\begin{equation}\label{eq.2}
M_{D} = \int_0^{r} \int_{-H/2}^{+H/2} 2\pi \rho \overline{r} d\overline{r} dz \sim \rho H r^{2\Delta} ,
\end{equation}
quite obviously, under $ \Delta \rightarrow 1 $ limit, the fractal medium will transform to a continuous medium.

\paragraph*{}
For a fractional infinitesimal volume element $ d\overline{V} = H \overline{r} d\overline{r} d\phi $ of the disc, the balance between mass flux and temporal change of density gives, the Continuity equation, of the form (see Roy \& Ray 2009\cite{Roy et al. 2009}),
\begin{equation}\label{eq.4}
\frac{\partial \Sigma }{\partial t} + \frac{1}{r^{2\Delta-1}} \frac{\partial}{\partial r} (\Sigma v r^{2\Delta-1}) = 0 .
\end{equation}
where, the  flow variables, $ \Sigma $ is the local surface density, and $ v $ is the radial drift velocity. In thin disc approximation $ \Sigma $, is defined by $ \Sigma \simeq \rho H $ (Frank et al. 2002\cite{Frank et al. 2002}).

\paragraph*{}
Similarly, the equation for radial momentum balance (i.e. Euler's equation), in the flow have to be constructed under the condition of fractal nature. This will finally lead to the result (see Roy \& Ray 2009\cite{Roy et al. 2009}), 
\begin{equation}\label{eq.10}
\frac{\partial v}{\partial t} + v \frac{\partial v}{\partial r} + \frac{1}{\rho} \frac{\partial P}{\partial r} + \phi'(r) -\frac{\lambda^2}{r^3} = 0 ,
\end{equation}
where, the pressure $ P $, can be expressed as a function of $ \rho $; $ \phi (r) $ is the generalised post-Newtonian pseudo-Kerr potential driving the flow (with the prime denoting a spatial derivative); $ \lambda $ is the constant specific angular momentum (Chakrabarti 1989\cite{Chakrabarti 1989}, 1990\cite{Chakrabarti 1990}, 1996\cite{Chakrabarti 1996}).

\paragraph*{}
As expected, only the continuity equation \eqref{eq.4} gets affected by the fractal nature of the flow medium, but, euler equation \eqref{eq.10} remains unchanged. The dynamics of velocity field $ v(r,t) $ and density field $ \rho(r,t) $, along with pressure $ P $(function of $ \rho $), describe the accretion flow of fractal matter, through equations \eqref{eq.4} and \eqref{eq.10}.

\paragraph*{}
The local thickness of the disc is $ H \equiv H(r) $. In fixing the function, $ H(r) $, one needs to look at the relevant geometrical configuration associated with accretion disc structures. This can vary in many ways, with different degrees of complexity (Chakrabarti \& Das 2001\cite{Chakrabarti et al. 2001}). In the simplest case one could treat $ H(r) $ to be just a constant, i.e. the disc is of uniform thickness (Constant Height Model, abbreviated as CH). In the case of the conical flow (Abramowicz \& Zurek 1981\cite{Abramowicz et al. 1981}) one prescribes, $ H(r) \propto r $ (Conical Model, abbreviated as CO). While these two cases could be viewed as giving an explicit dependence of $ H $ on $ r $, another well-invoked, but much more complicated prescription in accretion literature is that of the disc with the condition of hydrostatic equilibrium imposed in the vertical direction (Matsumoto et al. 1984\cite{Matsumoto et al. 1984}; Frank et al. 2002\cite{Frank et al. 2002}). In this particular instance, the function $ H(r) $ will be determined according to the way $ P $ has been prescribed (Frank et al. 2002\cite{Frank et al. 2002}). In all of these cases, however, it is a common practice to standardise transonicity in the flow by scaling its bulk velocity with the help of the local speed of sound  $ c_{s} = {\left( {{dP}/{d\rho}} \right) }^{1/2}  $ . In what follows, the equilibrium properties of the flow will be studied for the three different kinds of disc geometry mentioned above, under both polytropic and isothermal prescriptions for the equation of state.

The local thickness of the disk may be summarized for three flow geometries as follows. In constant height disk model (CH), the half thickness $H$ is independent of radial distance, i.e.,
\begin{equation}\label{eq.14}
H(r)_{CH} \cong H_{0} , \textrm{ where  $H_{0}$ \ is \ constant .} 
\end{equation}
But in the quasi-spherical or conical flow model (CO), it bears a linear dependence on $r$ as,
\begin{equation}\label{eq.15}
H(r)_{CO} \cong \Theta r , \textrm{ where \ $\Theta$ \ is \ constant .} 
\end{equation}\  
Finally the disk model (VE) where the disk thickness is determined by the hydrostatic equilibrium condition along axial direction, the half thickness is determined by the relation, 
\begin{equation}\label{eq.16}
H(r)_{VE} \cong c_{s} \left( \frac{r}{\gamma \phi'} \right)^{1/2} , \ where \ \phi \equiv \phi (r) . 
\end{equation}
in which, $ \phi (r) $ is the generalised post-Newtonian pseudo-Kerr potential driving the flow (with the prime denoting a spatial derivative).
\paragraph*{}
It is to be noted that the exponent of $ r $ in different disc height models and hence in mass continuity equation, differs by unity in CO and CH flow geometries while it is some complicated function in VE model. Thus, if one studies variation of $ \Delta $ (i.e. fractal nature of accreted matter) in one of these models (say in CO geometry), it will be identical to variation of $ \Delta $ in some other range in other models. Hence to get a qualitative estimate of the role of $ \Delta $, it is sufficient to focus on a single model. Here we focused on CO geometry for the sake of mathematical simplicity as well as reproducibility in time dependent numerical simulation. 

\section{STATIONARY FLOW OF FRACTAL MATTER}

\paragraph*{}
Under stationary condition $ \left(  \frac{\partial}{\partial t} = 0 \right) $, the two governing equations \eqref{eq.4} and \eqref{eq.10}, determining radial drift, will have the form, \\
the equation of continuity, 
\begin{equation}\label{eq.11}
\frac{\partial}{\partial r} \left(  \rho H(r) v r^{2\Delta-1} \right)  = 0 . 
\end{equation}
and the Euler's equation,
\begin{equation}\label{eq.12}
v \frac{\partial v}{\partial r} + \frac{1}{\rho} \frac{\partial P}{\partial r} + \phi'(r) -\frac{\lambda^2}{r^3} = 0.
\end{equation}

\paragraph*{}
The pressure, $ P $, being expressed in terms of density, $ \rho $, a polytropic equation of state (Chandrasekhar 1939\cite{Chandrasekhar 1939}), 
\begin{equation}\label{eq.13}
P=K\rho^\gamma ,
\end{equation}

where as the isothermal flow will be governed by the equation $ P={\rho \kappa_{B} T}/{\mu m_{H}} $. The quantities, $ K, \gamma, \kappa_{B}, T, \mu $ and $ m_{H} $ are the entropy per particle, the polytropic exponent, the Boltzmann's constant, the isothermal flow temperature, the reduced mass and the mass of a hydrogen atom, respectively. This $ \gamma $, allows us to treat both approximately adiabatic ($ \gamma \cong 5/3 $) and isothermal ($ \gamma \cong 1 $) accretion simultaneously. After the solution has been found, the adiabatic or isothermal assumption should be justified by consideration of the particular radiative cooling and heating of the gas. For example, the adiabatic approximation will be valid if the timescales for significant heating and cooling of the gas are long compared with the time taken for an element of the gas to fall in. In reality, neither extreme is quite satisfied, so we expect $ 1 \lesssim \gamma \lesssim 5/3 $ (see \cite{Juhan Frank Andrew King and Derek Raine Accretion Power in Astrophysics}).

\subsection{Polytropic Flows}

\paragraph*{}
With the help of $ H(r) $ for three different flow geometries, the integral solution of equation \eqref{eq.11} could be found out and we obtain the mass accretion rates for three different flow geometries :
\begin{equation}\label{eq.17}
\dot{M}_{CH} = \rho H_{0} v r^{2\Delta-1} ,
\end{equation}
\begin{equation}\label{eq.18}
\dot{M}_{CO} = \rho \Theta v r^{2\Delta} ,
\end{equation}
\begin{equation}\label{eq.19}
\dot{M}_{VE} = \rho c_{s}\left( \frac{1}{\gamma \phi'} \right)^{1/2}  v r^\sigma ,
\end{equation}
where  $ \sigma = 2\Delta-1/2 $ . \\
The corresponding entropy accretion rates can be obtained as
\begin{equation}\label{eq.20}
\mathcal{\dot{M}} = \dot{M} \left( \gamma K \right)^n , 
\end{equation}
where, $ n=\frac{1}{\gamma-1} $ . \\
Any function of $ K $, when multiplied by the total amount of mass flowing in per unit time, provides a measure of the total amount of inward entropy flux per unit time. $ \mathcal{\dot{M}} $ is thus called the entropy accretion rate. The concept of the entropy accretion rate was first introduced by Blaes(1987)\cite{Blaes 1987}.
\paragraph*{}
Using $ P=K\rho^\gamma $, ( hence $ {dP}/{d\rho} = c_{s}^2 = K\gamma\rho^{\gamma-1} $, and therefore $ \rho = \left( {c_{s}^2}/{\gamma K} \right)^n  $, with $ n={(\gamma-1)}^{-1} $ ),  the relation between the space gradient of the dynamical velocity and that of the sound speed may be obtained by differentiating the expressions for the corresponding entropy accretion rates for three different flow geometries.
\begin{equation}\label{eq.21}
\left( \frac{dc_{s}}{dr}\right)_{CH} = (1-\gamma) \frac{c_{s}}{2v} \left( \frac{dv}{dr}+(2\Delta-1) \frac{v}{r} \right) .
\end{equation}
\begin{equation}\label{eq.22}
\left( \frac{dc_{s}}{dr}\right)_{CO} = (1-\gamma) \frac{c_{s}}{v} \left( \frac{1}{2}\frac{dv}{dr}+ \Delta \frac{v}{r} \right) .
\end{equation}
\begin{equation}\label{eq.23}
\left( \frac{dc_{s}}{dr}\right)_{VE} = \left( \frac{1-\gamma}{1+\gamma} \right)  \frac{c_{s}}{v} \left[ \frac{dv}{dr} + \frac{v}{2} \left( \frac{2\sigma}{r}-\frac{\phi''(r)}{\phi'(r)} \right)  \right] .
\end{equation}

\paragraph*{}
The integral solution of the Euler equation provides the expression for the energy, first integral of motion, $\mathcal{E}$, also called Bernoulli's constant.
\begin{equation}\label{eq.24}
\frac{v^2}{2} + n c_{s}^2 + \frac{\lambda^2}{2r^2} + \phi(r) = \mathcal{E}.
\end{equation}

\paragraph*{}
We obtain the velocity space gradient by differentiating the algebraic expression for $ \mathcal{E} $ and by substituting the corresponding values of $ {d c_{s}}/{d r} $,
\begin{equation}\label{eq.25}
\left( \frac{d v}{d r} \right)_{CH} = \frac{\left[ \frac{\lambda^2}{r^3} + (2\Delta-1) \frac{c_{s}^2}{r} - \phi'(r) \right]}{\left( v - \frac{c_{s}^2}{v} \right)} . 
\end{equation}
\begin{equation}\label{eq.26}
\left( \frac{d v}{d r} \right)_{CO} = \frac{\left[ \frac{\lambda^2}{r^3} + 2\Delta \frac{c_{s}^2}{r} - \phi'(r) \right]}{\left( v - \frac{c_{s}^2}{v} \right)} . 
\end{equation}
\begin{equation}\label{eq.27}
\left( \frac{d v}{d r} \right)_{VE} = \frac{\left[ \frac{\lambda^2}{r^3} + \frac{\beta^2 c_{s}^2}{2} \left( \frac{2\sigma}{r} - \frac{\phi''(r)}{\phi'(r)} \right) - \phi'(r) \right]}{\left( v - \frac{\beta^2 c_{s}^2}{v} \right)} . 
\end{equation}
with $ \beta^2 = \frac{2}{(\gamma+1)} $. 

\subsubsection{Critical Points}

\paragraph*{}
Following the usual procedure adopted in the literature (see e.g. Nag et al. 2012\cite{Nag et al. 2012}; Chaudhury et al. 2006\cite{Chaudhury et al. 2006}) the critical point conditions can be obtained as
\begin{equation}\label{eq.28}
v_{c}^2 = c_{sc}^2 = \frac{1}{(2\Delta-1)} \left[ r_{c} \phi'(r_{c}) - \frac{\lambda^2}{r_{c}^2} \right],
\end{equation} for constant height flow.
\begin{equation}\label{eq.29}
v_{c}^2 = c_{sc}^2 = \frac{1}{(2\Delta)} \left[ r_{c} \phi'(r_{c}) - \frac{\lambda^2}{r_{c}^2} \right] ,
\end{equation}
for conical flow, and
\begin{equation}\label{eq.30}
v_{c}^2 = \beta^2 c_{sc}^2 = 2 \left[ r_{c} \phi'(r_{c}) - \frac{\lambda^2}{r_{c}^2} \right] \left[ 2\sigma - r_{c} \frac{\phi''(r_{c})}{\phi'(r_{c})} \right]^{-1} ,
\end{equation} for flow in vertical equilibrium; \\ 
where subscript `$ c $' stands for critical point values.
\paragraph*{}
Substitution of these critical point conditions (equations \eqref{eq.28},\eqref{eq.29},\eqref{eq.30}) into the expression for $ \mathcal{E} $ (equation \eqref{eq.24}) will provide a generalised algebraic form for the critical points, expressed in terms of the flow parameters, 
\begin{equation}\label{eq.31}
\frac{1}{2 (2\Delta-1)} \left( \frac{\gamma+1}{\gamma-1} \right) \left[ r_{c} \phi'(r_{c}) - \frac{\lambda^2}{r_{c}^2} \right] + \phi(r_{c}) + \frac{\lambda^2}{2 r_{c}^2} = \mathcal{E}  \ .
\end{equation} 
for constant height flow.
\begin{equation}\label{eq.32}
\frac{1}{4\Delta} \left( \frac{\gamma+1}{\gamma-1} \right) \left[ r_{c} \phi'(r_{c}) - \frac{\lambda^2}{r_{c}^2} \right] + \phi(r_{c}) + \frac{\lambda^2}{2 r_{c}^2} = \mathcal{E}  \ .
\end{equation} 
for conical flow.
\begin{equation}\label{eq.33}
\frac{2\gamma}{\gamma-1}\left[ r_{c} \phi'(r_{c}) - \frac{\lambda^2}{r_{c}^2} \right] \left[ 2\sigma - r_{c} \frac{\phi''(r_{c})}{\phi'(r_{c})} \right]^{-1} + \phi(r_{c}) + \frac{\lambda^2}{2 r_{c}^2} = \mathcal{E}  \ .
\end{equation} 
for flow in vertical equilibrium.
\paragraph*{}
With the help of these relations, it will then possible to fix the roots of $ r_{c} $ in terms of $ \gamma,\lambda,\Delta,\mathcal{E} $ and $ a $(`$ a $' will left it's imprint through the explicit functional form of $ \phi $).

\subsubsection{Nature Of The Critical Points: A Dynamical Systems Study}

\paragraph*{}
The governing fluid equations, describing the inviscide axisymmetric accretion flow onto a rotating black hole, belong to the general category of first-order nonlinear differential equations (Jordan \& Smith 1999\cite{Jordan et al. 1999}). There is no completely rigorous analytical prescription for solving these differential equations. A numerical integration is in most cases the only way for understanding the behaviour of the flow solutions. Alternatively another approach could be made to this question, if the governing equations are set up to form a standard first order dynamical system (Strogatz 1994\cite{Strogatz 1994}; Jordan \& Smith 1999\cite{Jordan et al. 1999}). This kind of approach is quite common in fluid dynamics (Bohr et al. 1993\cite{Bohr et al. 1993}), specially in accretion contex this kind of method has been successfully used before in several works (Ray \& Bhattacharjee 2002\cite{Ray et al. 2002}; Afshordi \& Paczy\'{n}ski 2003\cite{Afshordi et al. 2003}; Chaudhury et al. 2006\cite{Chaudhury et al. 2006}; Ray \& Bhattacharjee 2007\cite{Ray et al. 2007}; Mandal et al. 2007\cite{Mandal et al. 2007}; Goswami et al. 2007\cite{Goswami et al. 2007}, Nag et al. 2012\cite{Nag et al. 2012}). On doing so, it will first be necessary to parametrise the equations of stationary polytropic flow (equations \eqref{eq.25},\eqref{eq.26} and \eqref{eq.27}) and set up a coupled autonomous first-order dynamical system as (Strogatz 1994\cite{Strogatz 1994}; Jordan \& Smith 1999\cite{Jordan et al. 1999}), \\ 
for constant height disk model,
\begin{equation}\label{eq.34}
\frac{d (v^2)}{d \tau} = 2 v^2 \left[ \frac{\lambda^2}{r^2} - r \phi'(r) + (2\Delta-1) c_{s}^2  \right] ,
\end{equation}
\begin{equation}\label{eq.35}
\frac{d (r)}{d \tau} = r ( v^2 - c_{s}^2 ) .
\end{equation}
for conical flow model,
\begin{equation}\label{eq.36}
\frac{d (v^2)}{d \tau} = 2 v^2 \left[ \frac{\lambda^2}{r^2} - r \phi'(r) + (2\Delta) c_{s}^2  \right] ,
\end{equation}
\begin{equation}\label{eq.37}
\frac{d (r)}{d \tau} = r ( v^2 - c_{s}^2 ) .
\end{equation}
for flow under vertical hydrostatic equilibrium,
\begin{equation}\label{eq.38}
\frac{d (v^2)}{d \tau} = 2 v^2 \left[ \frac{\lambda^2}{r^2} - r \phi'(r) + \frac{\beta^2 c_{s}^2}{2} \left( 2\sigma-r\frac{\phi''(r)}{\phi'(r)} \right)  \right] ,
\end{equation}
\begin{equation}\label{eq.39}
\frac{d (r)}{d \tau} = r ( v^2 - \beta^2 c_{s}^2 ) .
\end{equation}
in which $ \tau $ is an arbitrary mathematical parameter. 

\paragraph*{}
Upon using a linear perturbation scheme about the critical points, as, $ v^2 = v_{c}^2 + \delta v^2 $, $ c_{s}^2 = c_{sc}^2 + \delta c_{s}^2 $ and $ r = r_{c} + \delta r $, it is possible to get a set of two autonomous first-order linear differential equations in $ \delta r - \delta v^2 $ plane. On doing so, it is necessary to first express $ \delta c_{s}^2 $ in terms of $ \delta r $ and $ \delta v^2 $, as (using equations \eqref{eq.21},\eqref{eq.22},\eqref{eq.23}), \\ 
in the case of constant height disk,
\begin{equation}\label{eq.40}
\frac{\delta c_{s}^2}{c_{s}^2} = - (\gamma-1) \left[ \frac{1}{2} \frac{\delta v^2}{v_{c}^2} + \left( 2 \Delta - 1 \right) \frac{\delta r}{r_{c}} \right] .
\end{equation}
For conical flow model,
\begin{equation}\label{eq.41}
\frac{\delta c_{s}^2}{c_{s}^2} = - (\gamma-1) \left[ \frac{1}{2} \frac{\delta v^2}{v_{c}^2} + \left( 2 \Delta \right) \frac{\delta r}{r_{c}} \right] .
\end{equation}
For the disk under vertical equilibrium condition,
\begin{equation}\label{eq.42}
\frac{\delta c_{s}^2}{c_{s}^2} = - \frac{\gamma-1}{\gamma+1} \left[ \frac{\delta v^2}{v_{c}^2} + \left( 2\sigma-r_{c}\frac{\phi''(r_{c})}{\phi'(r_{c})} \right) \frac{\delta r}{r_{c}} \right] .
\end{equation}
Finally, it is a straightforward exercise (see e.g. Goswami et al. 2007\cite{Goswami et al. 2007}, Bhattacharjee et al. 2009\cite{Bhattacharjee et al. 2009}), to establish the coupled linear dynamical system in the perturbed quantities $ \delta r $ and $ \delta v^2 $, as, \\
\begin{equation}\label{eq.43}
\frac{d}{d \tau} (\delta v^2) = A \delta v^2 + B \delta r ,
\end{equation}
\begin{equation}\label{eq.44}
\frac{d}{d \tau} (\delta r) = C \delta v^2 + D \delta r ,
\end{equation}
in which the constant coefficients $ A, B, C $ and $ D $ are to be read as \\ 

For constant height disc flow,
\begin{equation}\label{eq.45}
A = - (2\Delta-1) (\gamma-1) c_{sc}^2 \ ,
\end{equation}
\begin{eqnarray}\nonumber\label{eq.46}
B & = & -2 c_{sc}^2 \Biggl[ \frac{2\lambda^2}{r_{c}^3} + \phi'(r_{c})  
\\
&& + r_{c} \phi''(r_{c}) + (2\Delta-1)^2 (\gamma-1) \frac{c_{sc}^2}{r_{c}}  \Biggr] \ ,
\end{eqnarray}
\begin{equation}\label{eq.47}
C = \left( \frac{\gamma+1}{2} \right) r_{c} \ ,
\end{equation}
\begin{equation}\label{eq.48}
D = (2\Delta-1) (\gamma-1) c_{sc}^2 \ .
\end{equation}

For conical flow,
\begin{equation}\label{eq.49}
A = -2 \Delta (\gamma-1) c_{sc}^2 \ ,
\end{equation}
\begin{eqnarray}\nonumber\label{eq.50}
B & = & -2 c_{sc}^2 \Biggl[ \frac{2\lambda^2}{r_{c}^3} + \phi'(r_{c}) 
\\
&& + r_{c} \phi''(r_{c}) +  4 \Delta^2 (\gamma-1) \frac{c_{sc}^2}{r_{c}}  \Biggr] \ ,
\end{eqnarray}
\begin{equation}\label{eq.51}
C = \left( \frac{\gamma+1}{2} \right) r_{c} \ ,
\end{equation}
\begin{equation}\label{eq.52}
D = 2\Delta (\gamma-1) c_{sc}^2 \ .
\end{equation}

For the flow under vertical hydrostatic equilibrium,
\begin{equation}\label{eq.53}
A = \left( \frac{\gamma-1}{\gamma+1} \right) \mathcal{X} v_{c}^2 \ ,
\end{equation}
\begin{eqnarray}\nonumber\label{eq.54}
B & = & -2 v_{c}^2 \Biggl[ \frac{2\lambda^2}{r_{c}^3} + \phi'(r_{c}) 
\\ \nonumber 
&& + r_{c} \phi''(r_{c}) + \frac{1}{2} \beta^2 c_{sc}^2 \frac{\phi''(r_{c})}{\phi'(r_{c})} \mathcal{Y} 
\\
&& + \frac{1}{2} \frac{\beta^2 c_{sc}^2}{r_{c}} \left( \frac{\gamma-1}{\gamma+1} \right) \mathcal{X}^2  \Biggr] \ ,
\end{eqnarray}
\begin{equation}\label{eq.55}
C = \left( \frac{2\gamma}{\gamma+1} \right) r_{c} \ ,
\end{equation}
\begin{equation}\label{eq.56}
D = - \left( \frac{\gamma-1}{\gamma+1} \right) v_{c}^2 \mathcal{X} \ ,
\end{equation}
under the further definition that
\begin{equation}\label{eq.57}
\mathcal{X} = r_{c} \frac{\phi''(r_{c})}{\phi'(r_{c})} -2\sigma \ ,
\end{equation}
\begin{equation}\label{eq.58}
\mathcal{Y} = 1 + r_{c} \frac{\phi'''(r_{c})}{\phi''(r_{c})} -r_{c} \frac{\phi''(r_{c})}{\phi'(r_{c})} \ .
\end{equation}

\paragraph*{}
Trial solutions of the type $ \delta v^2 \ \sim \exp(\Omega\tau) $ and $ \delta r \ \sim \exp(\Omega\tau) $ in equations \eqref{eq.43},\eqref{eq.44} will provide the eigenvalues $ \Omega $, which are the growth rates of $ \delta v^2 $ and $ \delta r $, as
\begin{equation}\label{eq.59}
\Omega^2-(A+D)\Omega+(AD-BC)=0 .
\end{equation}
Under a further definition that $ P=A+D $ , $ Q = AD - BC $ and $ \Xi=P^2-4Q $ , the solution of the foregoing quadratic equation can be written as
\begin{equation}\label{eq.60}
\Omega= \frac{P\pm \sqrt{\Xi}}{2} .
\end{equation}
\paragraph*{}
Once the numerical value of $r_c $ is known, the nature of the critical point can be obtained by studying the values of $\Omega^2$.
\paragraph*{}
The nature of the possible critical points can also be predicted from the form of $ \Omega $ in equation \eqref{eq.60}. If $ \Xi > 0 $, then a critical point can be either a saddle or a node (Jordan \& Smith 1999\cite{Jordan et al. 1999}). The precise nature of the critical point will then be dependent on the sign of $ Q $. If $ Q < 0 $, then the critical point will be a saddle point. Such points are always notoriously unstable in terms of the sensitivity in generating a solution through them, after starting from a boundary value far away from the critical point (Ray \& Bhattacharjee 2002\cite{Ray et al. 2002}, 2007\cite{Ray et al. 2007}; Roy \& Ray 2007\cite{Roy et al. 2007}). On the other hand, if $ Q > 0 $, then the critical point will be a node. Such a point may or may not be stable, depending on the sign of $ P $. If $ P<0 $, then the node will be stable.
\paragraph*{}
A completely different class of critical points will result when $ \Xi<0 $. These points will be like a spiral (a focus). Once again, the stability of the spiral will depend on the sign of $ P $. If $ P<0 $, then the spiral will be stable. Which will obviously mean that if the critical point is either a spiral or a node, then it will be stable, with flow solutions in the neighborhood of the critical point converging towards it.
\paragraph*{}
Noting that a centre-type point $ (P = 0) $ is merely a special case of a spiral, and then for a centre-type point $ P = A+D = 0 $ , therefore from equation \eqref{eq.60},
\begin{equation}\label{eq.61}
\Omega^2 = B C - A D = - Q .
\end{equation}
\paragraph*{}
For an inviscid fractal disc flow, the allowed critical points will be either saddle points or centre-type points. If, $ \Omega^2 >0 $, then the critical point will be a saddle point, on the other hand if $ \Omega^2 <0 $, then it will be a centre-type point, with $ \Omega^2 $ having real values for both of the cases. 

\subsubsection{Slope Of The Continuous Solutions Passing Through The Critical Points}
\paragraph*{}
If the critical points are known, transonic accretion solutions can be obtained by integrating the corresponding expressions for $ {dv}/{dr} $ subjected to the critical value of the $ {dv}/{dr} $, i.e., the value of $ {dv}/{dr} $ evaluated at the critical point(s). The slope of the continuous solutions which could possibly pass through the critical points are to be obtained by applying the L'Hospital rule on equations \eqref{eq.25},\eqref{eq.26},\eqref{eq.27} and using equations \eqref{eq.21},\eqref{eq.22},\eqref{eq.23} at the critical points. This will give a quadratic equation for the slope of the stationary solutions at the critical points.  The resulting expression i.e. the values of $ {dv}/{dr} $ evaluated at the critical point(s),will read as, \\ 
for constant height disk geometry,
\begin{eqnarray}\nonumber\label{eq.62}
\frac{dv}{dr}\Biggr|_{r_{c}} & = & \ \ \left( \frac{1-\gamma}{1+\gamma} \right) \frac{\sqrt{2\Delta-1}}{r_{c}} \sqrt{ \left( r_{c} \phi' - \frac{\lambda^2}{r_{c}^2} \right) }
\\ \nonumber 
&& \pm \Biggl[ \left( \frac{1-\gamma}{1+\gamma} \right)^{2} \frac{2 \Delta-1}{r_{c}^{2}} \left( r_{c} \phi' - \frac{\lambda^2}{r_{c}^2} \right)  
\\ \nonumber  
&& - \frac{1}{\gamma+1} \biggl\lbrace   \frac{(2\Delta-1) (\gamma-1) +1}{r_{c}^{2}} \left( r_{c} \phi' - \frac{\lambda^2}{r_{c}^2} \right) 
\\
&& + \frac{3 \lambda^{2} }{r_{c}^4} + \phi'' \biggr\rbrace \Biggr] ^{1/2} .
\end{eqnarray}

For conical disk geometry,
\begin{eqnarray}\nonumber\label{eq.63}
\frac{dv}{dr}\Biggr|_{r_{c}} & = & \ \ \left( \frac{1-\gamma}{1+\gamma} \right) \frac{2 \Delta}{r_{c}} \sqrt{ \frac{1}{2 \Delta} \left( r_{c} \phi' - \frac{\lambda^2}{r_{c}^2} \right) }
\\ \nonumber 
&& \pm \Biggl[ \left( \frac{1-\gamma}{1+\gamma} \right)^{2} \frac{2 \Delta}{r_{c}^{2}} \left( r_{c} \phi' - \frac{\lambda^2}{r_{c}^2} \right)  
\\ \nonumber 
&& - \frac{1}{\gamma+1} \biggl\lbrace \frac{2 \Delta (\gamma-1) +1}{r_{c}^{2}} \left( r_{c} \phi' - \frac{\lambda^2}{r_{c}^2} \right)
\\
&& + \frac{3 \lambda^{2} }{r_{c}^4} + \phi'' \biggr\rbrace \Biggr] ^{1/2} .
\end{eqnarray}

For vertical equilibrium disk geometry,
\begin{eqnarray}\nonumber\label{eq.64}
\frac{dv}{dr}\Biggr|_{r_{c}} & = & \ \ \frac{\gamma-1}{2\gamma v_{c}} \left( \frac{\lambda^2}{r_{c}^3} - \phi' \right) 
\\ \nonumber 
&& \pm \frac{\gamma+1}{4\gamma v_{c}} \Biggl[ 4\frac{(\gamma-1)^2}{(\gamma+1)^2} \left( \phi'-\frac{\lambda^2}{r_{c}^3} \right)^2 
\\ \nonumber 
&& -\frac{4 \gamma v_{c}^2}{\gamma+1} \biggl\lbrace \frac{\gamma-1}{\gamma+1} \left( \phi'-\frac{\lambda^2}{r_{c}^3} \right) \left( \frac{2\sigma}{r_{c}}-\frac{\phi''}{\phi'} \right) 
\\ \nonumber 
&& +\frac{v_{c}^2}{2}\left( \frac{2\sigma}{r_{c}^2}-\frac{\phi''^2}{\phi'^2}+\frac{\phi'''}{\phi'} \right)
\\
&& + \frac{3 \lambda^{2} }{r_{c}^4} + \phi'' \biggr\rbrace \Biggr] ^{1/2} .
\end{eqnarray}

\subsubsection{Numerical Results}
\paragraph*{}
A set of values of $ [\mathcal{E},\gamma,\lambda,\Delta,a] $ is required to solve the algebraic expressions and hence to obtain the value of the corresponding critical point $ r_{c} $. One usually uses the range $[0\lesssim\mathcal{E}<1, 4/3\leqslant\gamma\leqslant5/3, 0<\lambda<4, 0.5 \leqslant \Delta \leqslant 1, -1 \leqslant a \leqslant 1]$, e.g., see Saha et al. 2016\cite{Saha et al. 2016}. 
\paragraph*{}

\begin{figure}[H]
\centering
\includegraphics[angle=0,width=0.5\textwidth]{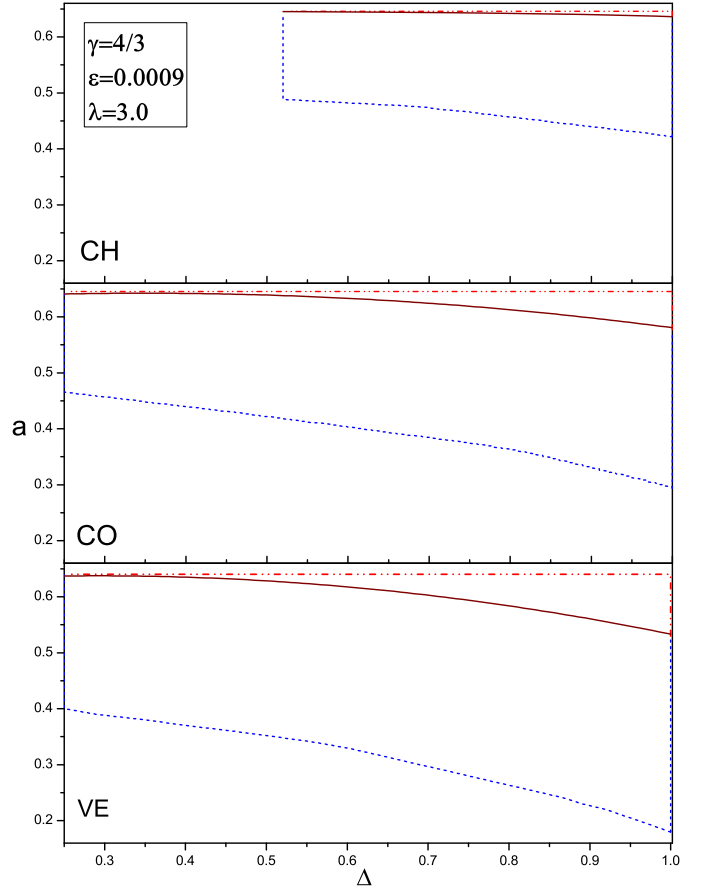}
\caption{For polytropic flows, region of multitransonicity in the parameter space of $ a $ and $ \Delta $, with $ \gamma=4/3,\mathcal{E}=0.0009, \lambda=3.0 $ for different disc geometry models (see text).}\label{param_a_del_aw_chcove}
\end{figure}


\paragraph*{}
Considering polytropic flows, \autoref{param_a_del_aw_chcove} shows the plot of $ a $ and $ \Delta $ multicritical parameter space for different disc geometry models, CH, CO and VE respectively, subjected to the pseudo-Newtonian potential used here. For each $ a - \Delta $ parameter space, the region bounded by the dashed lines (Blue) and solid lines (Brown) is characterised by the condition $\dot{\mathcal{M}}_{in}>\dot{\mathcal{M}}_{out}$ where $\dot{\mathcal{M}}_{in}$ and $\dot{\mathcal{M}}_{out}$ are the entropy accretion rate corresponding to the stationary integral flow solution passing through the inner and the outer critical points respectively; the region bounded by the dashed double dot lines (Red) and solid lines (Brown) is characterised by the criteria $\dot{\mathcal{M}}_{in}<\dot{\mathcal{M}}_{out}$. 

\begin{figure}[H]
\centering
\includegraphics[angle=0,width=0.5\textwidth]{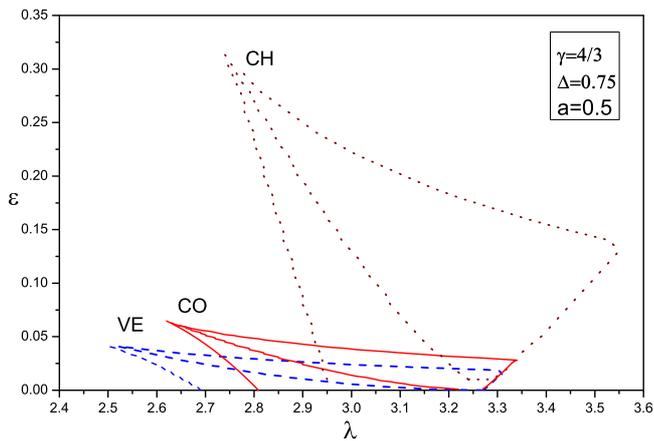}
\caption{For polytropic flows, region of multitransonicity in the parameter space of $ \mathcal{E} $ and $ \lambda $, with $ \gamma=4/3,a=0.5,\Delta=0.75 $. The dotted lines (Brown) are for CH flow, the solid lines (Red) are for CO flow, and the small dashed lines (Blue) are for VE flow. For notations see text.}\label{f1}
\end{figure}

\paragraph*{}
\autoref{f1} gives the multicritical parameter space of $ \mathcal{E} $ and $ \lambda $, in different disk geometry models for CH, CO and VE respectively. For all three geometries, there are certain wedge shaped regions which correspond to three critical points and outside these regions, the parameter values shown in the figure, generate single critical point only. For each model, inside the parameter space, left portion of the region is characterised  by the condition $\dot{\mathcal{M}}_{in}>\dot{\mathcal{M}}_{out}$ and right portion is characterised by $\dot{\mathcal{M}}_{in}<\dot{\mathcal{M}}_{out}$.

\begin{figure}[H]
\centering
\includegraphics[angle=0,width=0.5\textwidth]{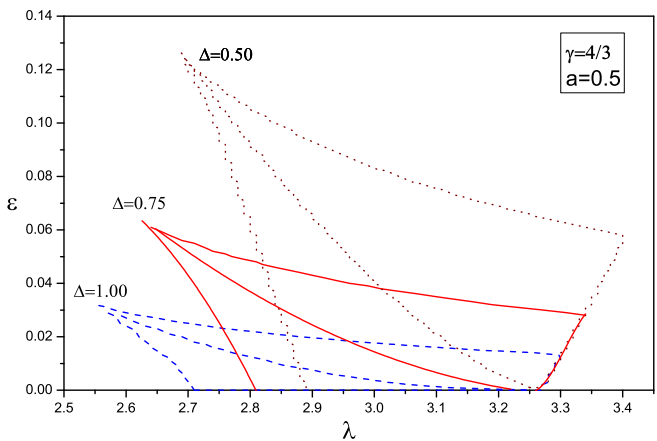}
\caption{ For CO polytropic flow, region of multitransonicity in the parameter space of $ \mathcal{E} $ and $ \lambda $ , with $ \gamma=4/3,a=0.5 $ and for varying $ \Delta $ (Dotted line (Brown) corresponds to $ \Delta=0.5 $, Solid line (Red) corresponds to $ \Delta=0.75 $, Dashed line (Blue) corresponds to $ \Delta=1.0 $).}\label{f2}
\end{figure}

\begin{figure}[H]
\centering
\includegraphics[angle=0,width=0.5\textwidth]{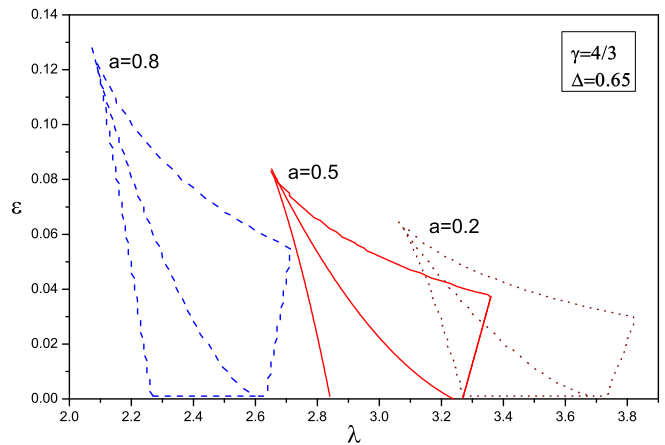}
\caption{ For CO polytropic flow, region of multitransonicity in the parameter space of $ \mathcal{E} $ and $ \lambda $ , with $ \gamma=4/3,\Delta=0.65 $ and for varying $ a $ (Dotted line (Brown) corresponds to $ a=0.2 $, Solid line (Red) corresponds to $ a=0.5 $, Dashed line (Blue) corresponds to $ a=0.8 $).}\label{f3}
\end{figure}

\paragraph*{}
It should now be instructive to consider how the variation of spin parameter $ a $ and fractal parameter $ \Delta $ affect the critical properties, because for the case of a rotating black hole in a fractal medium, this two parameters will also leave its imprint on the physics of the accretion process. The area of multitransonicity shifts according to the choice of $ a $ and $ \Delta $. \autoref{f2} shows that for a fixed value of $ \gamma $ and $ a $ the $ \mathcal{E}-\lambda $ parameter space shifts according to the variation of $ \Delta $. \autoref{f3} shows that for a fixed value of $ \gamma $ and $ \Delta $ the $ \mathcal{E}-\lambda $ parameter space shifts according to the variation of $ a $. For the increase in Fractal nature (decrease in $ \Delta $), the multitransonic region shifts towards higher values of both $ \mathcal{E} $ and $ \lambda $ and as well as the area of multitransonicity increases. In case of black hole spin parameter, when rotation of the black hole is more pronounced ($ a $ increases), the shift of the multitransonic region towards lower values of $ \lambda $ and higher values of $ \mathcal{E} $ is quite evident. Here also the area of multitransonicity increases with the increase in the value of $ a $.

\begin{figure}[H]
\centering
\includegraphics[angle=0,width=0.5\textwidth]{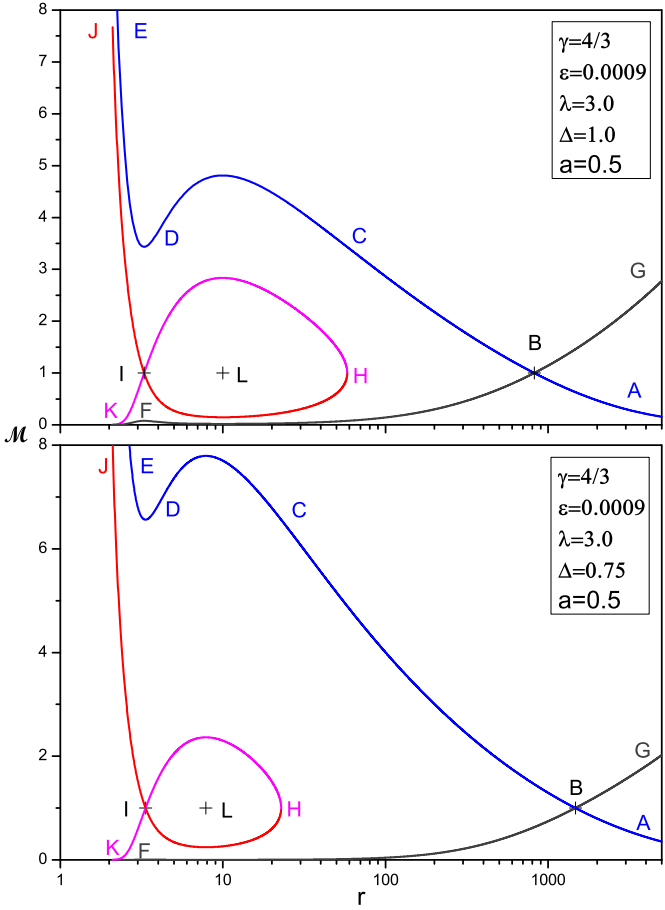}
\caption{ For CO geometry, phase portrait of polytropic transonic accretion in continuous ($ \Delta=1.0 $) and fractal medium ($ \Delta=0.75 $) respectively,  as a function of radius $ r $  for $ \gamma = 4/3 ,\ \mathcal{E} = 0.0009 ,\ \lambda=3.0$ and $\ a=0.5 $.}\label{phase_plots}
\end{figure}

\paragraph*{}
For conical polytropic flow, we have shown a representative phase portrait for multi-transonic accretion in both non fractal medium (with $ \Delta=1.0 $) and fractal medium (with $ \Delta=0.75 $) in  \autoref{phase_plots}. ABCDE (Blue line) is the transonic accretion branch through the outer sonic point B. The solution JIHIK (Red, Magenta line) through the inner sonic point does not connect the outer boundary of $r$ at infinity. The middle critical point is L. The other branch FBG (Gray line) in \autoref{phase_plots} is themwind solution and hence does not have much relevance in the context of accretion flow. Keeping all the other parameters fixed (at $ \gamma = 4/3 ,\ \mathcal{E} = 0.0009 ,\ \lambda=3.0 , \ a=0.5 $) the introduction of fractal nature via the change in fractal parameter ($ \Delta $), shows that the position of the outer critical point get shifted outward in fractal medium.

\begin{figure}[H]
\centering
\includegraphics[angle=0,width=0.5\textwidth]{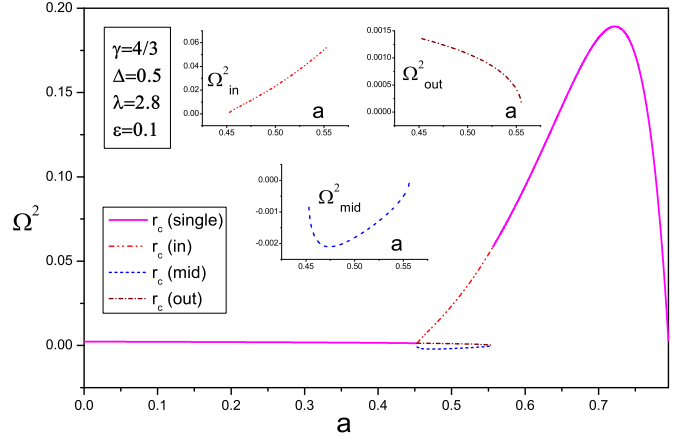}
\caption{For CO polytropic flow, the variation of $ \Omega^{2} $ with $ a $ (see text). Here the chosen parameter values are $ \gamma=4/3 $, $ \Delta=0.5 $, $ \lambda=2.8 $, $ \mathcal{E}=0.1 $.}\label{omega2_a}
\end{figure}

\begin{figure}[H]
\centering
\includegraphics[angle=0,width=0.5\textwidth]{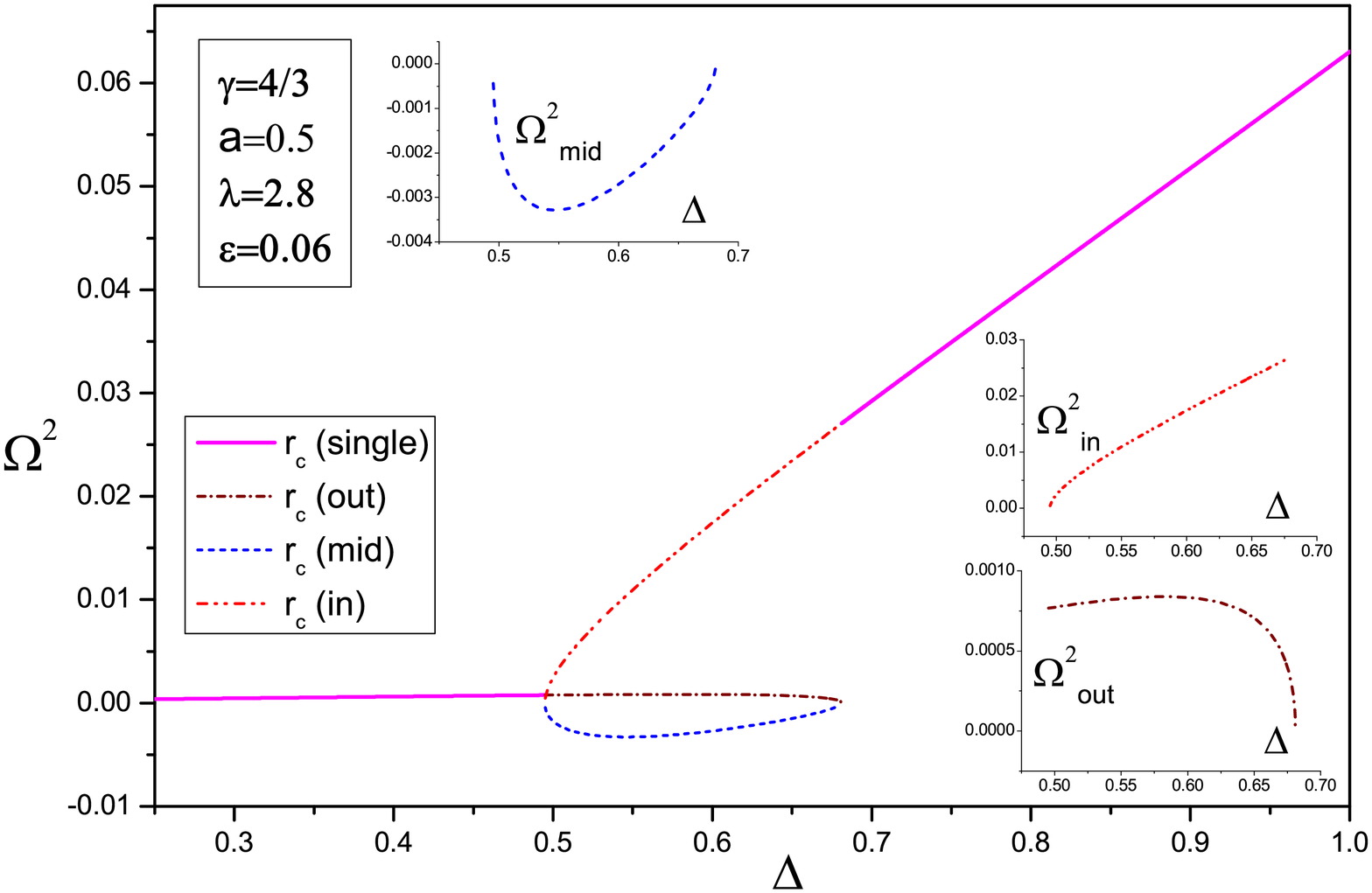}
\caption{For CO polytropic flow, the variation of $ \Omega^{2} $ with $ \Delta $ (see text). Here the chosen parameter values are $ \gamma=4/3 $, $ a=0.5 $, $ \lambda=2.8 $, $ \mathcal{E}=0.06 $.}\label{omega2_del}
\end{figure}

\paragraph*{}
To derive some quantitative insight about the specific properties of an individual critical point, however, it will be necessary to examine the behaviour of the eigenvalues of the stability matrix associated with each critical point. This has to be done by going back to \eqref{eq.61}, which gives a dependence of $ \Omega^{2} $ on the critical point coordinates. These coordinates, in their turn, have a dependence on the parameters $ [\mathcal{E},\gamma,\lambda,\Delta,a] $. Keeping the first three parameters fixed, the variation of $ \Omega^{2} $ with respect to spin parameter $ a $ and also with respect to fractal parameter $ \Delta $, have been plotted, for all critical points, in \autoref{omega2_a} and \autoref{omega2_del}, respectively. In both plots the solid lines (Magenta) are for single critical points, the dash double dot line (Red) is for inner critical points, the dash line (Blue) is for middle critical points and the dash dot line (Brown) is for outer critical points. 

\begin{figure}[H]
\centering
\includegraphics[angle=0,width=0.5\textwidth]{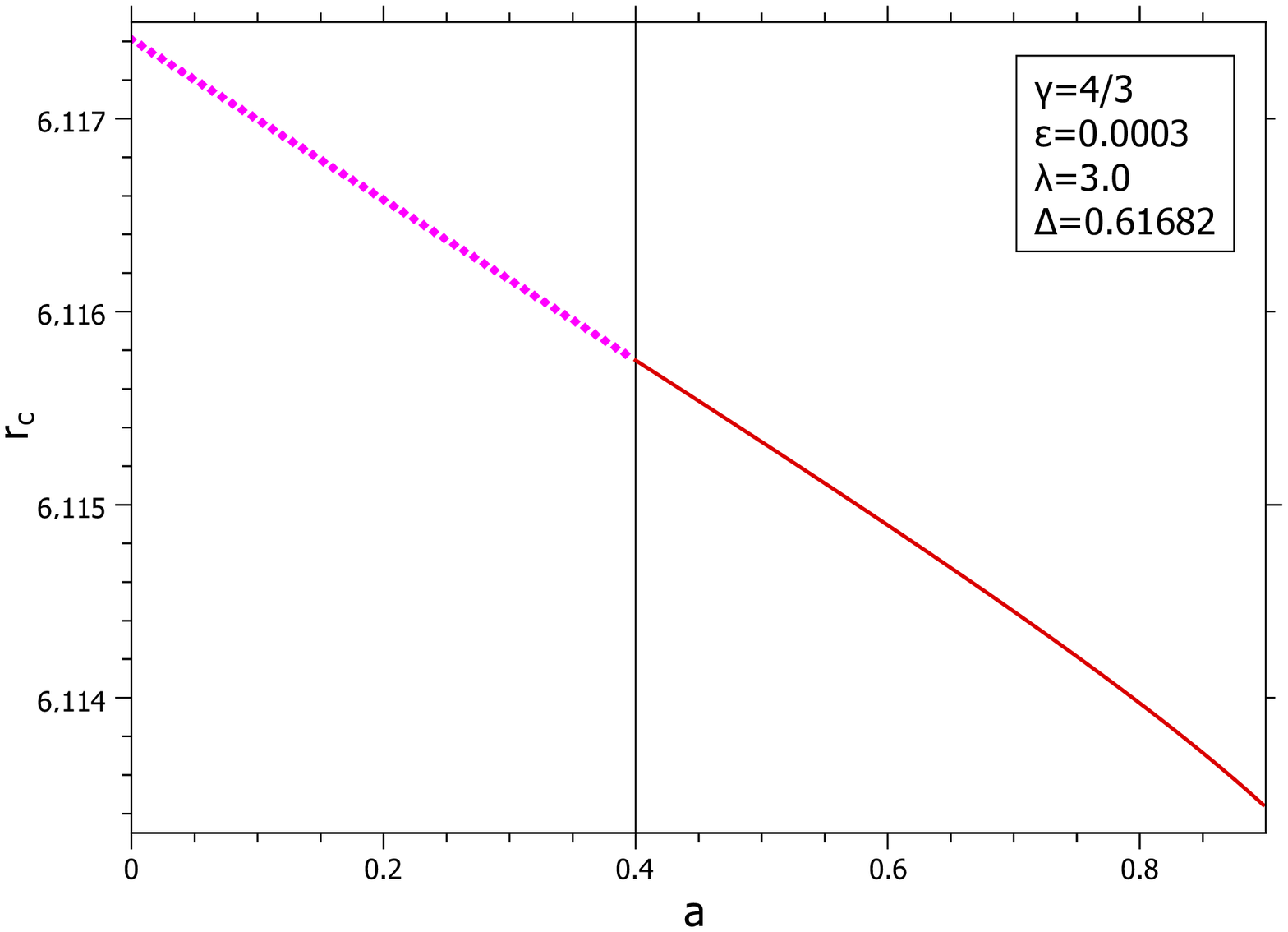}
\caption{For CO polytropic flow, the locus of the position of the single critical point, dotted (Magenta) line (left to the vertical line) and of the outermost critical point, solid (Brown) line (right to the vertical line), for varying $ a $. Here the chosen parameter values are $ \gamma=4/3 $, $ \mathcal{E}=0.0003 $, $ \lambda=3.0 $, $ \Delta=0.61682 $. The vercical (Black) line separets the Multitransonic region (right) and lone critical point region (left). }\label{f5}
\end{figure}

\begin{figure}[H]
\centering
\includegraphics[angle=0,width=0.5\textwidth]{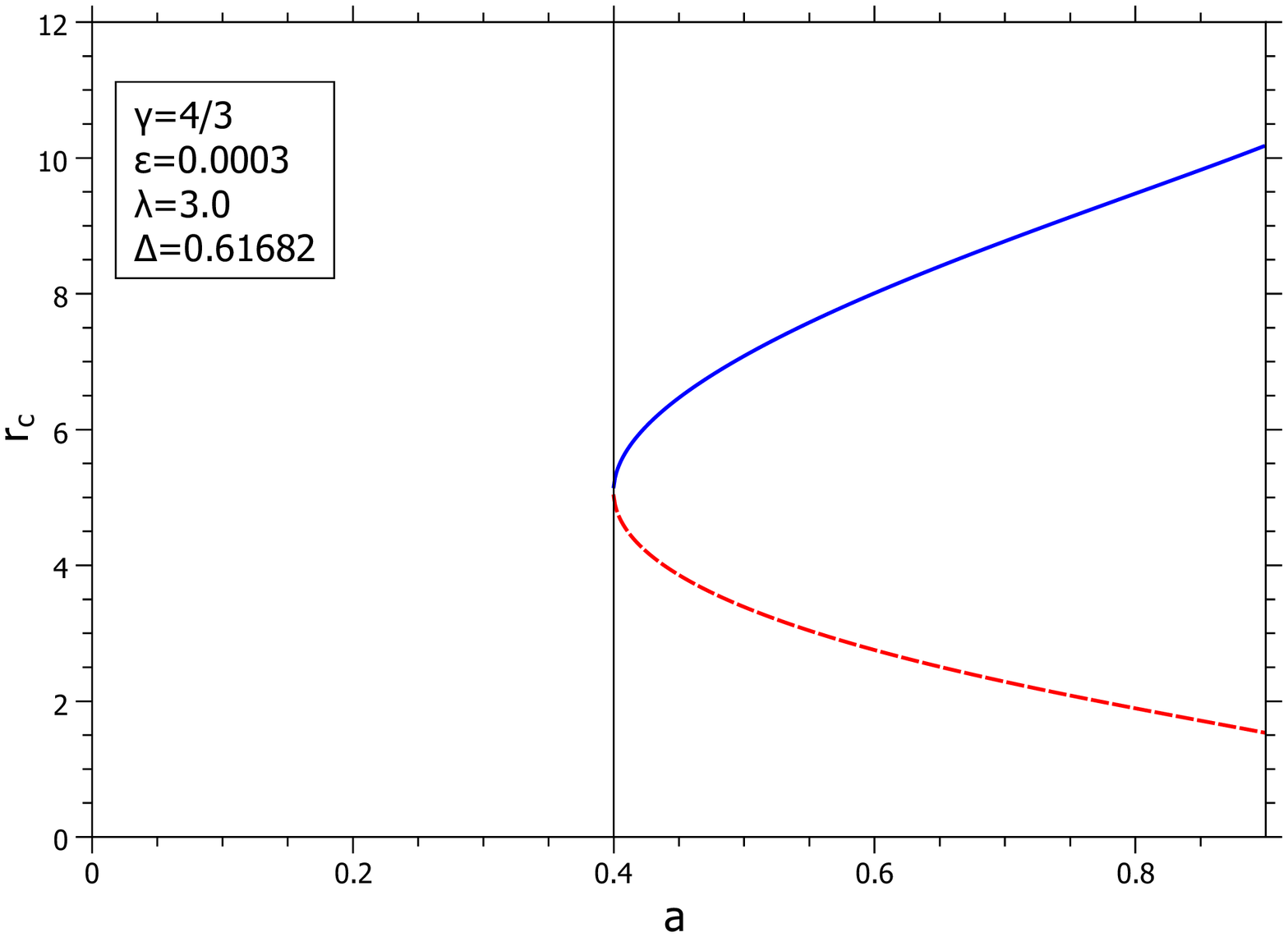}
\caption{ For CO polytropic flow, the locus of the position of the middle critical point (the upper arm of the cusp, solid (Blue) line) and innermost critical point (the lower arm of the cusp, dashed (Red) line), for varying $ a $. Here the chosen parameter values are $ \gamma=4/3 $, $ \mathcal{E}=0.0003 $, $ \lambda=3.0 $, $ \Delta=0.61682 $. The two tracks merges at ($ r_{c} \approx 5.09  $, when $ a=0.4 $) the vertical line (Black line), and thus it is boundary of the Multitransonic region (right).}\label{f6}
\end{figure}

\begin{figure}[H]
\centering
\includegraphics[angle=0,width=0.5\textwidth]{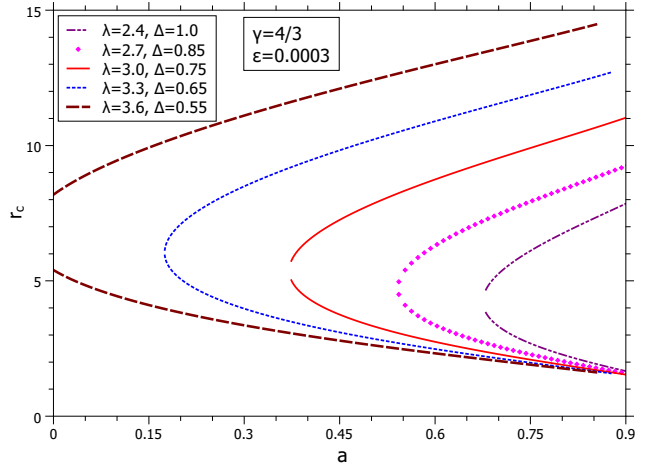}
\caption{For CO polytropic flow, the locus of the position of the middle critical point (the upper arm of the cusp) and innermost critical point (the lower arm of the cusp), with varying $ a $, for different values of $ \Delta $ and $ \lambda $ (see text). Here the chosen parameter values are $ \gamma=4/3 $, $ \mathcal{E}=0.0003 $.}\label{f7}
\end{figure}

\paragraph*{}
Surveying the whole range ($ 0 \leq a \leq 0.9 $) of spin parameter $ a $, the vertical line (Black) at $ a=0.4 $, in \autoref{f5} and \autoref{f6}, separates the critical points in monotransonic and multitransonic category. To the left of this line is the region having one root only (indicated by Dotted (Magenta) line in \autoref{f5}). On the right is the region having three real roots; outer most critical points (indicated by Solid (Brown) line in \autoref{f5}), middle critical points (indicated by Solid (Blue) arm in \autoref{f6}) and inner most critical points (indicated by Dashed (Red) arm in \autoref{f6}). In both of the region, there is a root, exists always, shown by \autoref{f5}. In \autoref{f6} the two roots (middle and inner critical points), making a cusp, annihilated mutually at the vertical line. For increasing $ a $, across the range $ 0 \leq a \leq 0.9 $, initially there is only one critical point, which exists forever and later there is a birth of other two critical points, simultaneously i.e. bifurcation occurs. So, multitransonicity is only possible within a limited range of spin parameter, not for all values. For simultaneous increase in $ \lambda $ and decrease in $ \Delta $, the vertical line, passing through the bifurcation point, shifts towards the left (in \autoref{f7}, double dot dashed cusp (Purple) is for $ \lambda=2.4,\Delta=1.0 $; long separated dot cusp (Magenta) is for $ \lambda=2.7,\Delta=0.85 $; solid cusp (Red) is for $ \lambda=3.0,\Delta=0.75 $; dotted cusp (Blue) is for $ \lambda=3.3,\Delta=0.65 $ and dashed cusp (Brown) is for $ \lambda=3.6,\Delta=0.55 $).

\begin{figure}[H]
\centering
\includegraphics[angle=0,width=0.5\textwidth]{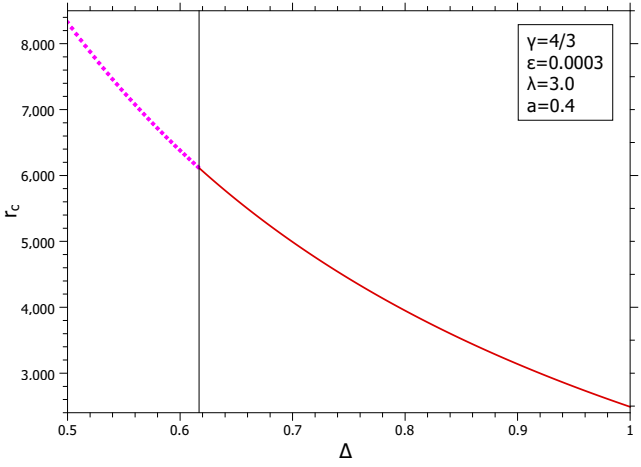}
\caption{For CO polytropic flow, the locus of the position of the single critical point, dotted (Magenta) line (left to the vertical line) and of the outermost critical point, solid (Brown) line (right to the vertical line), for varying $ \Delta $. Here the chosen parameter values are $ \gamma=4/3 $, $ \mathcal{E}=0.0003 $, $ \lambda=3.0 $, $ a=0.4 $. The vercical (Black) line separets the Multitransonic region (right) and lone critical point region (left).}\label{f8}
\end{figure}

\begin{figure}[H]
\centering
\includegraphics[angle=0,width=0.5\textwidth]{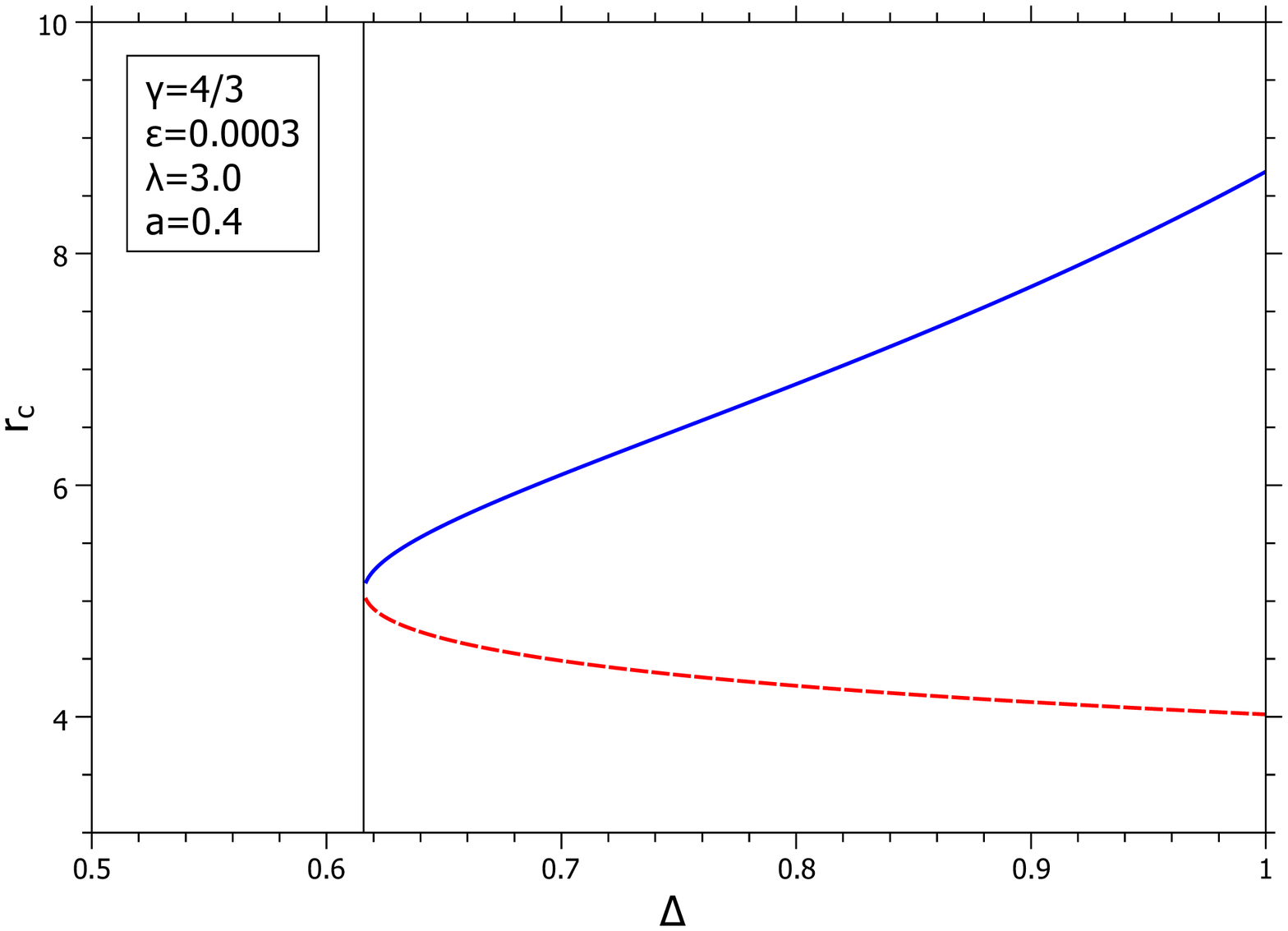}
\caption{ For CO polytropic flow, the locus of the position of the middle critical point (the upper arm of the cusp, solid (Blue) line) and innermost critical point (the lower arm of the cusp, dashed (Red) line), for varying $ \Delta $. Here the chosen parameter values are $ \gamma=4/3 $, $ \mathcal{E}=0.0003 $, $ \lambda=3.0 $, $ a=0.4 $. The two tracks merges at ($ r_{c} \approx 5.09  $, when $ \Delta=0.6169 $) the vertical line (Black line), and thus it is boundary of the Multitransonic region (right).}\label{f9}
\end{figure}

\begin{figure}[H]
\centering
\includegraphics[angle=0,width=0.5\textwidth]{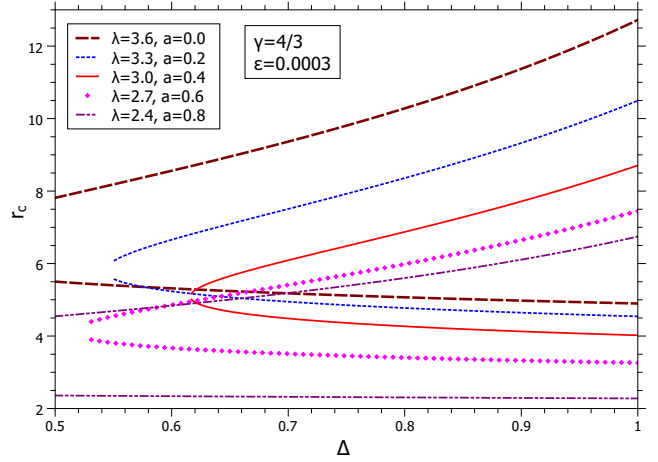}
\caption{For CO polytropic flow, the locus of the position of the middle critical point (the upper arm of the cusp) and innermost critical point (the lower arm of the cusp), with varying $ \Delta $, for different values of $ a $ and $ \lambda $ (see text). Here the chosen parameter values are $ \gamma=4/3 $, $ \mathcal{E}=0.0003 $.}\label{f10}
\end{figure}

\paragraph*{}
Surveying the whole range ($ 0.5 \leq \Delta \leq 1.0 $) of fractal parameter $ \Delta $, the vertical line (Black) at $ \Delta=0.6169 $, in \autoref{f8} and \autoref{f9}, separates the critical points in monotransonic and multitransonic category. To the left of this line is the region having one root only (indicated by Dotted (Magenta) line in \autoref{f8}). On the right is the region having three real roots; outer most critical points (indicated by Solid (Brown) line in \autoref{f8}), middle critical points (indicated by Solid (Blue) arm in \autoref{f9}) and inner most critical points (indicated by Dashed (Red) arm in \autoref{f9}). In both of the region, there is a root, exists always, shown by \autoref{f8}. In \autoref{f9}, the two roots (middle and inner critical points), making a cusp, annihilated mutually at the vertical line. For increasing $ \Delta $, across the range $ 0.5 \leq \Delta \leq 1.0 $, initially there is only one critical point, which exists forever and later there is a birth of other two critical points, simultaneously i.e. bifurcation occurs. So, multitransonicity is only possible within a limited range of fractal parameter, not for all values. For simultaneous increase in $ \lambda $ and decrease in $ a $, the vertical line, passing through the bifurcation point, shifts first towards right and then towards left (in \autoref{f10}, the double dot dashed cusp (Purple) is for $ \lambda=2.4,a=0.8 $; long separated dot cusp (Magenta) is for $ \lambda=2.7,a=0.6 $; solid cusp (Red) is for $ \lambda=3.0,a=0.4 $; dotted cusp (Blue) is for $ \lambda=3.3, a=0.2 $ and dashed cusp (Brown) is for $ \lambda=3.6, a=0.0 $).

\subsection{Isothermal Flows}

\paragraph*{}
For an isothermal flow, the full mathematical treatment is actually much simpler. Here one has to go back to equation \eqref{eq.12} and use the linear dependence between $ P $ and $ \rho $ using the appropriate equation of state. The integral solution of the time independent Euler equation gives,
\begin{equation}\label{eq.72}
\frac{v^2}{2} + c_{s}^2 \ln \rho + \frac{\lambda^2}{2r^2} + \phi(r) = \mathcal{C},
\end{equation}
in which, $ \mathcal{C} $ is a constant of integration. \\ 
The first integral solution of continuity equation gives, the mass accretion rates, for three disc geometry models, as, \\
for constant height flow,
\begin{equation}\label{eq.73}
 \rho H_{0} v r^{(2\Delta-1)}  = \dot{M}.
\end{equation}
for conical flow,
\begin{equation}\label{eq.74}
\rho \Theta v r^{2\Delta}  = \dot{M}.
\end{equation}
for flow under vertical equilibrium condition,
\begin{equation}\label{eq.75}
\frac{\rho c_{s} v r^{\sigma} }{ \sqrt{\phi'} } = \dot{M},
\end{equation}
where  $ \sigma = 2\Delta-1/2 $ . \\

\paragraph*{}
The global constant, isothermal sound speed, $ c_{s} $, can be expressed as, $ c_{s} = \varphi T^{1/2} $, with $ \varphi = \left( {\kappa_{B}}/{\mu m_{H}}\right)^{1/2} $ and $ T $ is the isotherm flow temperature. Using equations \eqref{eq.72} and \eqref{eq.73},\eqref{eq.74},\eqref{eq.75}, the space gradient of the velocities for these three models comes out to be,
\begin{equation}\label{eq.76}
\left( \frac{d v}{d r} \right)_{CH} = \frac{\left[ \frac{\lambda^2}{r^3} + (2\Delta-1) \frac{c_{s}^2}{r} - \phi'(r) \right]}{\left( v - \frac{c_{s}^2}{v} \right)} . 
\end{equation}
\begin{equation}\label{eq.77}
\left( \frac{d v}{d r} \right)_{CO} = \frac{\left[ \frac{\lambda^2}{r^3} + 2\Delta \frac{c_{s}^2}{r} - \phi'(r) \right]}{\left( v - \frac{c_{s}^2}{v} \right)} . 
\end{equation}

\begin{equation}\label{eq.78}
\left( \frac{d v}{d r} \right)_{VE} = \frac{\left[ \frac{\lambda^2}{r^3} + \frac{ c_{s}^2}{2} \left( \frac{2\sigma}{r} - \frac{\phi''(r)}{\phi'(r)} \right) - \phi'(r) \right]}{\left( v - \frac{ c_{s}^2}{v} \right)} . 
\end{equation}
which gives the critical point conditions as, \\
for constant height disk,
\begin{equation}\label{eq.79}
v_{c}^2 = c_{s}^2 = \frac{1}{(2\Delta-1)} \left[ r_{c} \phi'(r_{c}) - \frac{\lambda^2}{r_{c}^2} \right] .
\end{equation}
For conical flow model,
\begin{equation}\label{eq.80}
v_{c}^2 = c_{s}^2 = \frac{1}{(2\Delta)} \left[ r_{c} \phi'(r_{c}) - \frac{\lambda^2}{r_{c}^2} \right] .
\end{equation}
For the disk under vertical hydrostatic equilibrium condition,
\begin{equation}\label{eq.81}
v_{c}^2 = c_{s}^2 = 2 \left[ r_{c} \phi'(r_{c}) - \frac{\lambda^2}{r_{c}^2} \right] \left[ 2\sigma - r_{c} \frac{\phi''(r_{c})}{\phi'(r_{c})} \right]^{-1} .
\end{equation}
where the subscript $ c $ stands for the critical point values, as usual. Here, for the isothermal flow the critical points can be obtained by solving the equation of $ r_{c} $ in terms of flow parameters $ T, \lambda, \Delta, a $.

\begin{figure}[H]
\centering
\includegraphics[angle=0,width=0.5\textwidth]{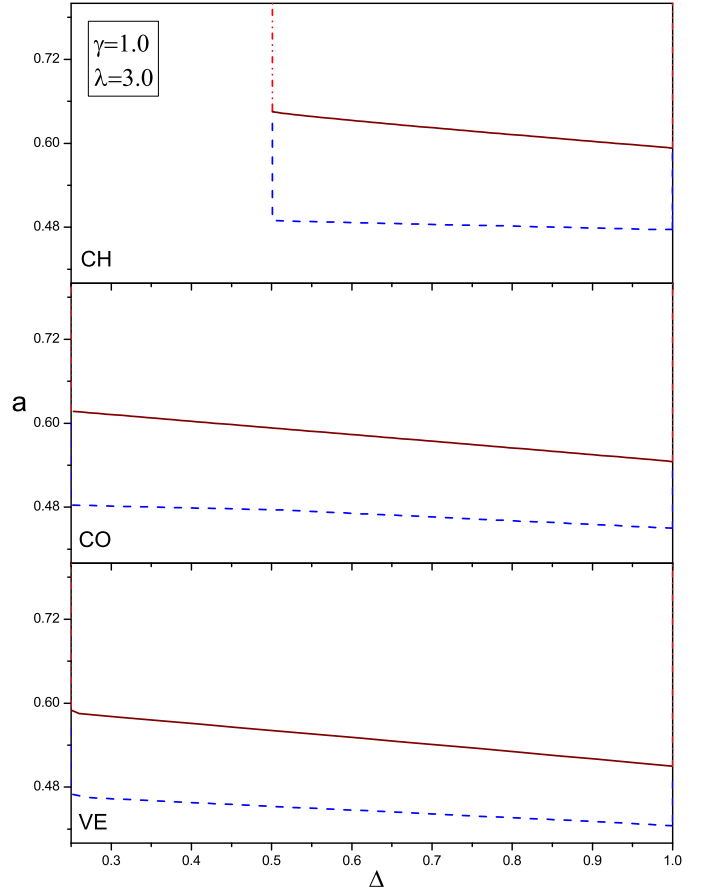}
\caption{For isothermal flows, region of multitransonicity in the parameter space of $ a $ and $ \Delta $, with $ \gamma=1.0,T=9.39\times10^{10} K, \lambda=3.0 $ for different disc geometry models (see text).}\label{param_a_del_aw_chcove_iso}
\end{figure}

\begin{figure}[H]
\centering
\includegraphics[angle=0,width=0.5\textwidth]{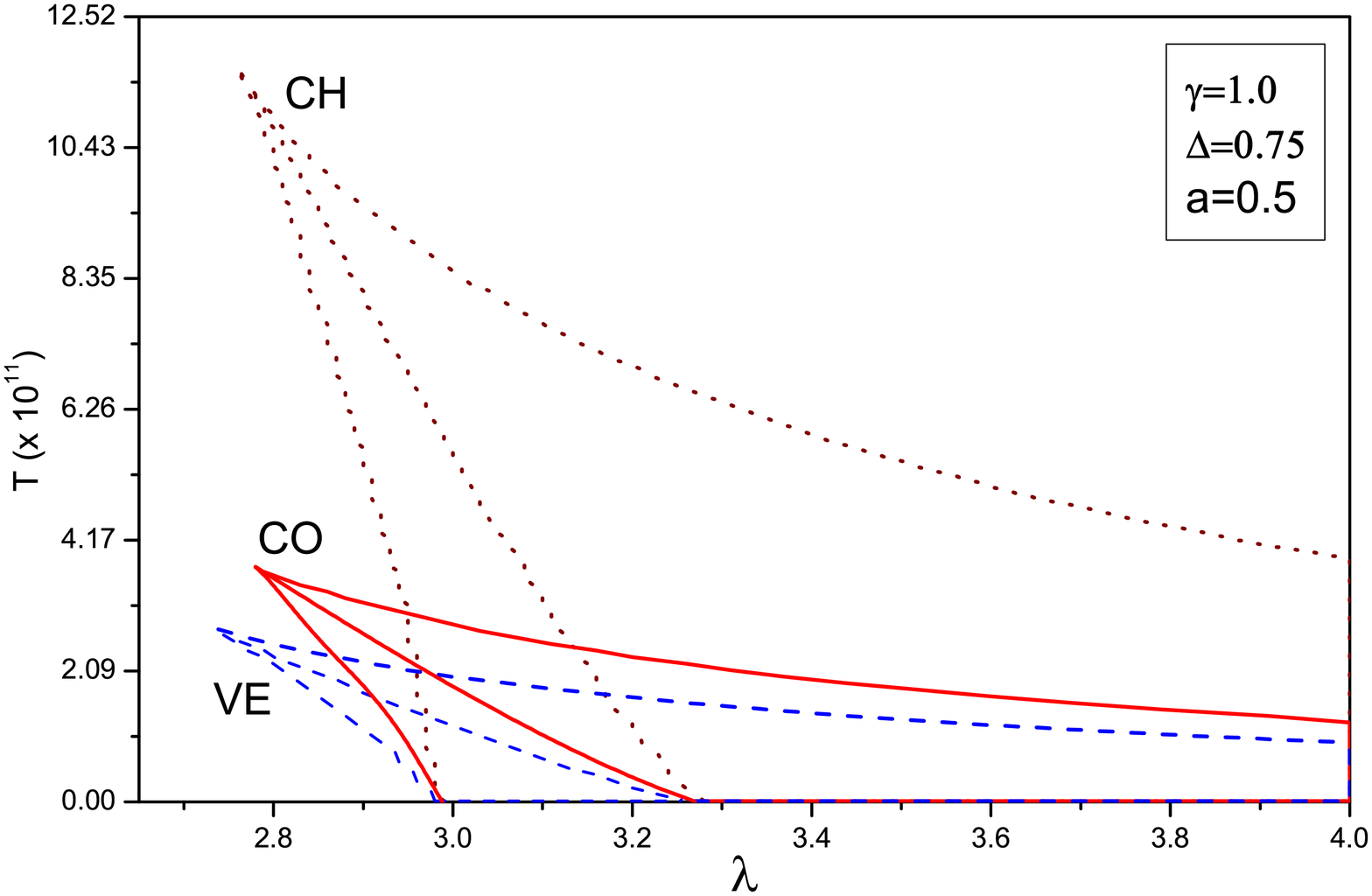}
\caption{For isothermal flows, region of multitransonicity in the parameter space of $ T $ and $ \lambda $, with $ \gamma=1.0,a=0.5,\Delta=0.75 $. The dotted lines (Brown) are for CH flow, the solid lines (Red) are for CO flow, and the small dashed lines (Blue) are for VE flow. For notations see text.}\label{param_T_l_chcove_iso}
\end{figure}

\begin{figure}[H]
\centering
\includegraphics[angle=0,width=0.5\textwidth]{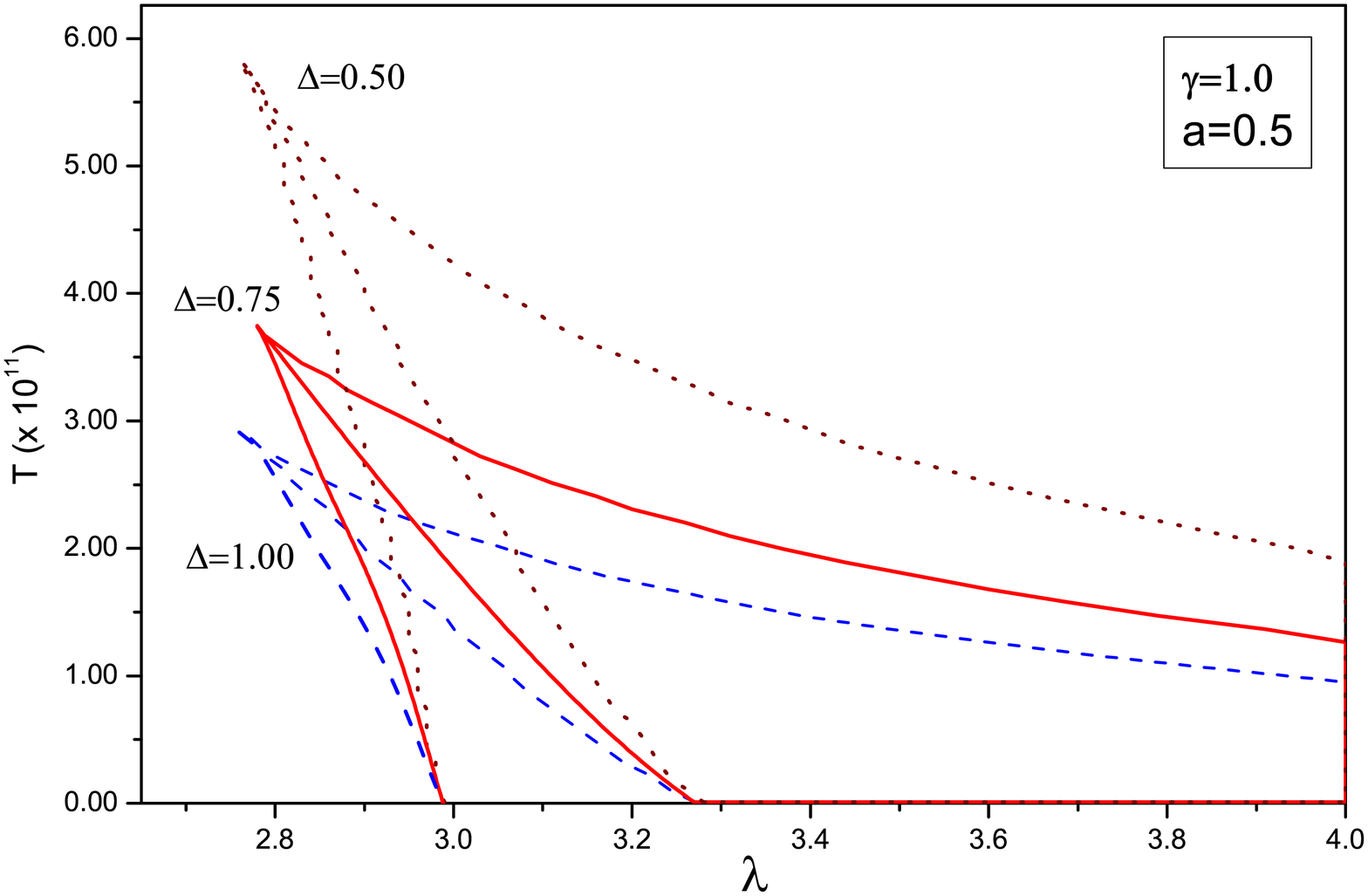}
\caption{ For CO isothermal flow, region of multitransonicity in the parameter space of $ T $ and $ \lambda $, with $ \gamma=1.0,a=0.5 $ and for varying $ \Delta $ (Dotted line (Brown) corresponds to $ \Delta=0.5 $, Solid line (Red) corresponds to $ \Delta=0.75 $, Dashed line (Blue) corresponds to $ \Delta=1.0 $).}\label{param_T_l_a0pt5_iso}
\end{figure}

\begin{figure}[H]
\centering
\includegraphics[angle=0,width=0.5\textwidth]{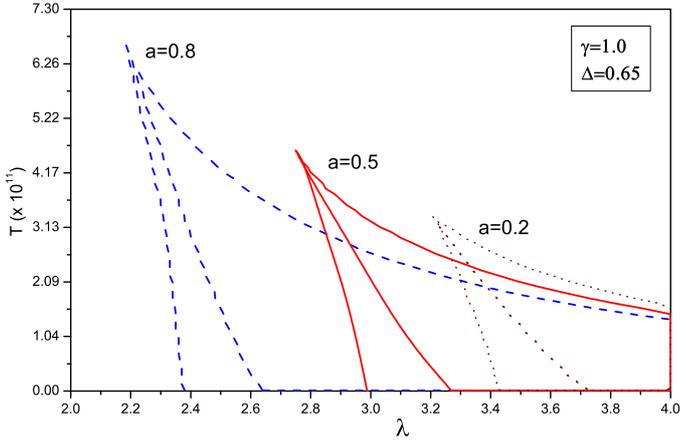}
\caption{ For CO isothermal flow, region of multitransonicity in the parameter space of $ T $ and $ \lambda $, with $ \gamma=1.0,\Delta=0.65 $ and for varying $ a $ (Dotted line (Brown) corresponds to $ a=0.2 $, Solid line (Red) corresponds to $ a=0.5 $, Dashed line (Blue) corresponds to $ a=0.8 $).}\label{param_T_l_del0pt65_iso}
\end{figure}

\paragraph*{}
With the use of isothermal equation of state, \autoref{param_a_del_aw_chcove_iso}, gives different regions of $ a-\Delta $ parameter space for which multitransonicity is observed, corresponding to different flow geometry models, say CH, CO and VE. For a fixed value of $ \left[ \gamma, a, \Delta \right] $, \autoref{param_T_l_chcove_iso} shows the plot of $ T-\lambda $ multicritical parameter space in different disk geometries. Considering CO model, for fixed values of other parameter, the change in $ T-\lambda $ parameter space due to variation of $ \Delta $ and $ a $ are depicted by \autoref{param_T_l_a0pt5_iso} and \autoref{param_T_l_del0pt65_iso}, respectively. As for the polytropic case, here the multicritical parameter space (\autoref{param_a_del_aw_chcove_iso}, \autoref{param_T_l_chcove_iso}, \autoref{param_T_l_a0pt5_iso} and \autoref{param_T_l_del0pt65_iso}), can also be divided into sub-regions, characterized by the quantity $ \mathcal{C} $. Although  all of these plots for isothermal flow, follow the same trend as for the polytropic flow, but, unlike polytropic case here multitransonicity can also be obtained even with higher values of spin parameter $ a $ ($ a-\Delta $ plots) or with higher values of specific angular momentum $ \lambda $ ($ T-\lambda $ plots), i.e. for very strongly rotating flows (denoted by $ \mathcal{C}_{in} > \mathcal{C}_{out} $ sub-region).

\begin{figure}[H]
\centering
\includegraphics[angle=0,width=0.5\textwidth]{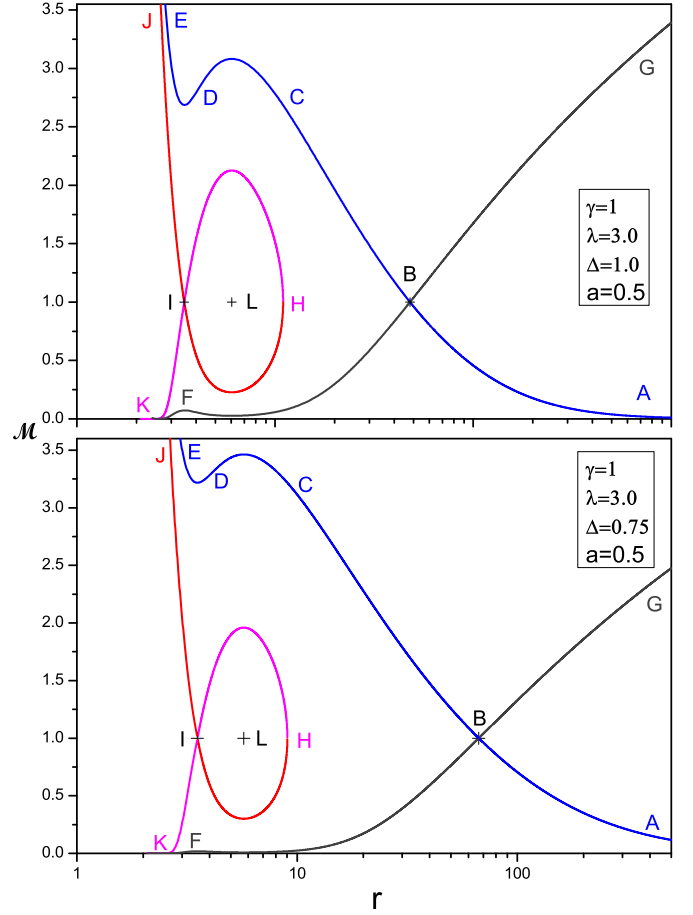}
\caption{ For CO geometry, phase portrait of isothermal transonic accretion, in continuous ($ \Delta=1.0 $) and fractal medium ($ \Delta=0.75 $) respectively, for $ \gamma = 1, T=9.39\times 10^{10}K,\ \lambda=3.0,\ \Delta=0.75, \ a=0.5 $.}\label{f11}
\end{figure}

\paragraph*{}
A typical multi-transonic conical flow topology is shown, for both non fractal medium (with $ \Delta=1.0 $) and fractal medium
(with $ \Delta=0.75 $), in \autoref{f11}, for $ [T=9.39\times 10^{10}K,\lambda=3.0,\Delta=0.75,a=0.5] $. All the plots for the isothermal flow, follow the same trend as obtained earlier, in case of polytropic flow.
\begin{figure}[H]
\centering
\includegraphics[angle=0,width=0.5\textwidth]{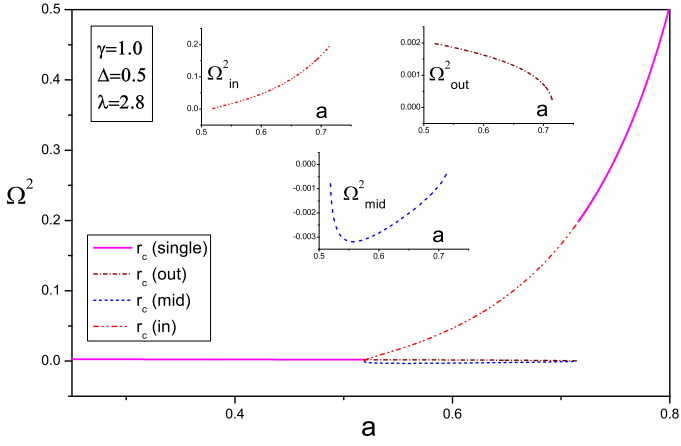}
\caption{For CO isothermal flow, the variation of $ \Omega^{2} $ with $ a $. Here the chosen parameter values are $ \gamma=1 $, $ \Delta=0.5 $, $ \lambda=2.8 $, $ T=4.38\times10^{11} K $.}\label{omega2_a_iso}
\end{figure}

\begin{figure}[H]
\centering
\includegraphics[angle=0,width=0.5\textwidth]{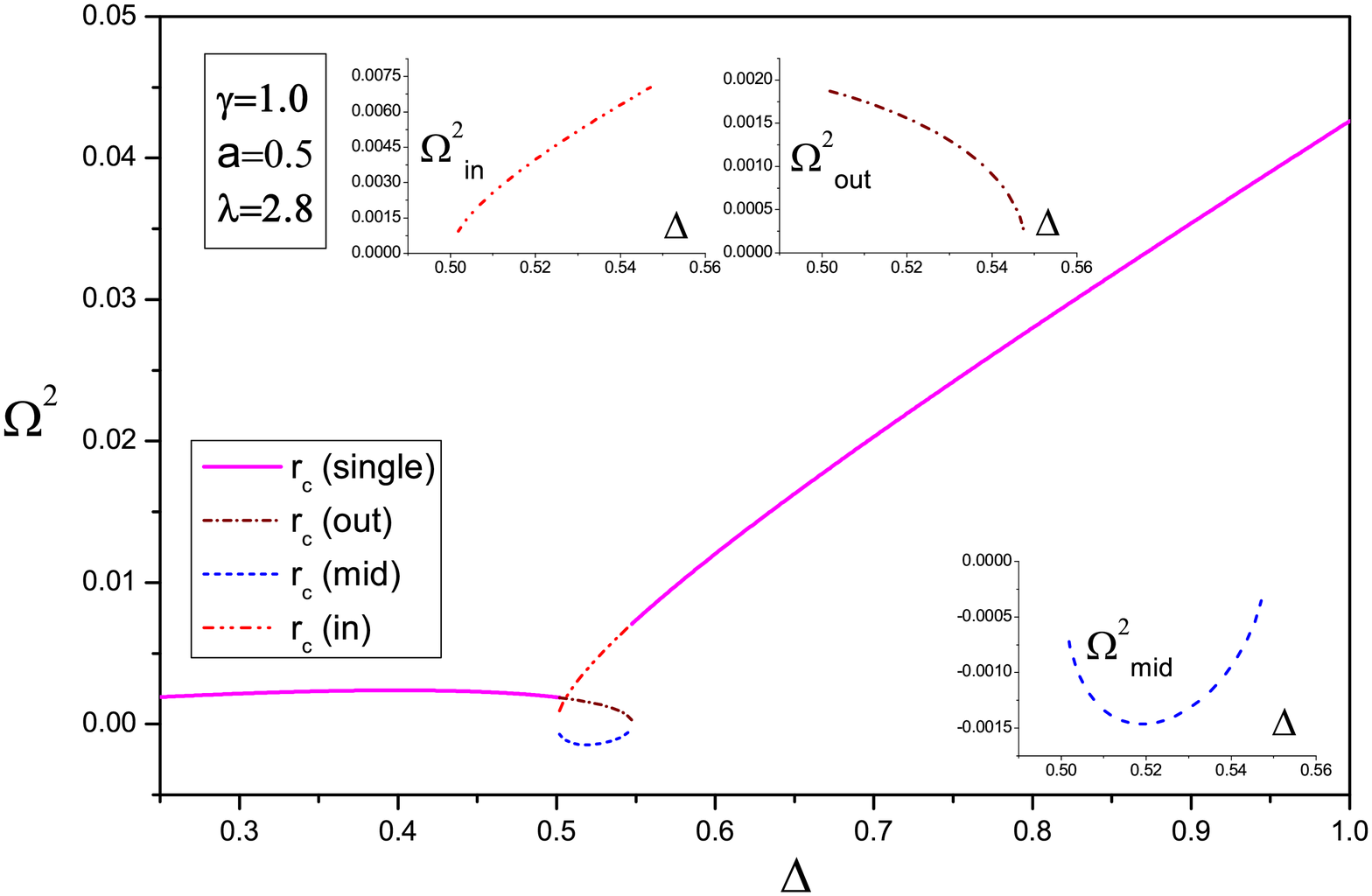}
\caption{ For CO isothermal flow, the variation of $ \Omega^{2} $ with $ \Delta $. Here the chosen parameter values are $ \gamma=1 $, $ a=0.5 $, $ \lambda=2.8 $, $ T=5.00\times10^{11} K $.}\label{omega2_del_iso}
\end{figure}

\begin{figure}[H]
\centering
\includegraphics[angle=0,width=0.5\textwidth]{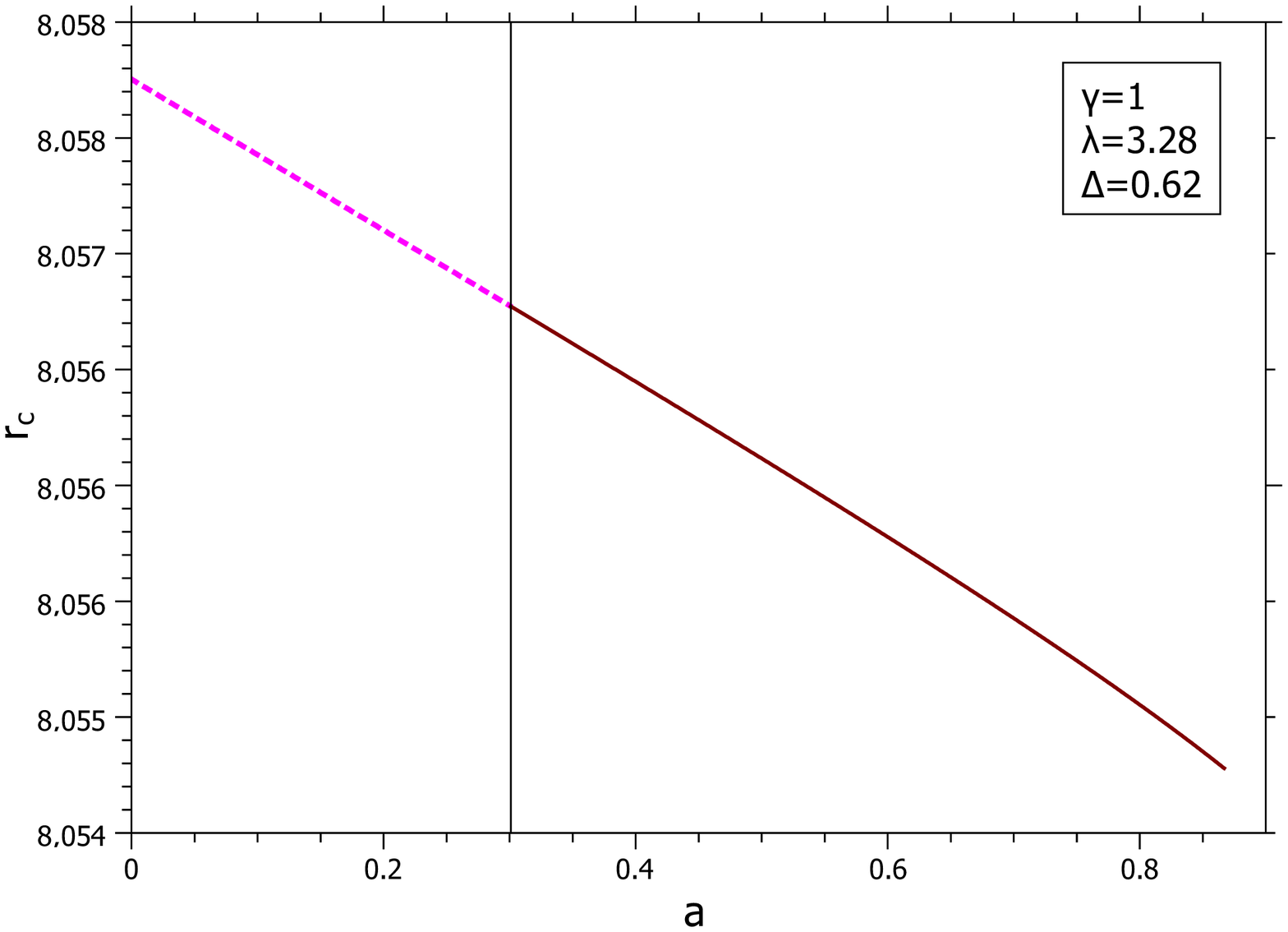}
\caption{For CO isotremal flow, the locus of the position of the single critical point, dotted (Magenta) line (left to the vertical line) and of the outermost critical point, solid (Brown) line (right to the vertical line), for varying $ a $. Here the chosen parameter values are $ \gamma=1 $, $ T \simeq 10^{11}K $, $ \lambda=3.28 $, $ \Delta=0.62 $. The vercical (Black) line separets the Multitransonic region (right) and lone critical point region (left). }\label{f12}
\end{figure}

\begin{figure}[H]
\centering
\includegraphics[angle=0,width=0.5\textwidth]{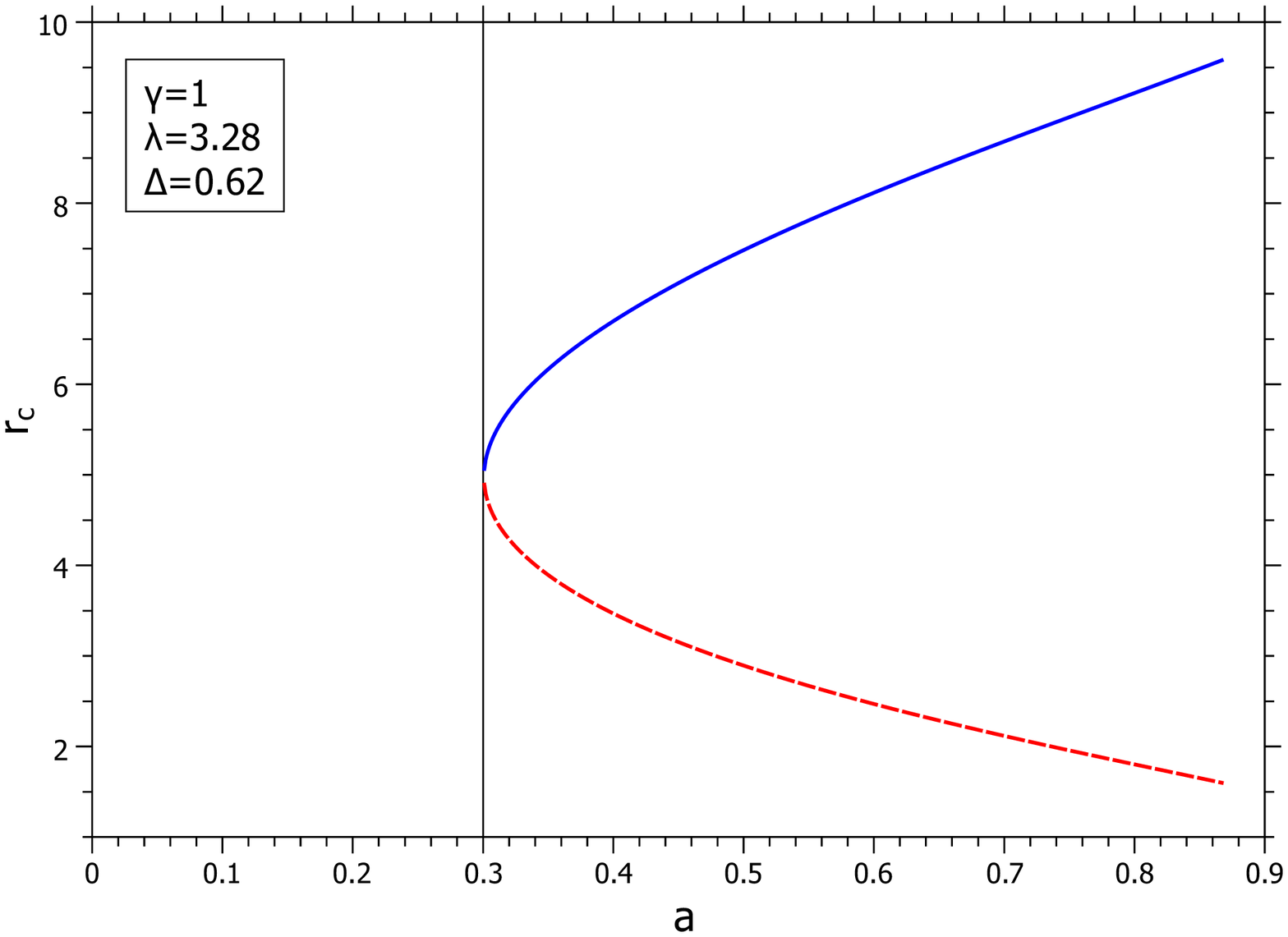}
\caption{For CO isothermal flow,  the locus of the position of the middle critical point (the upper arm of the cusp, solid (Blue) line) and innermost critical point (the lower arm of the cusp, dashed (Red) line), for varying $ a $. Here the chosen parameter values are $ \gamma=1 $, $ T \simeq 10^{11}K $, $ \lambda=3.28 $, $ \Delta=0.62 $. The two tracks merges at ($ r_{c} \approx 4.978  $, when $ a=0.301 $) the vertical line (Black line), and thus it is boundary of the Multitransonic region (right)}\label{f13}
\end{figure}

\begin{figure}[H]
\centering
\includegraphics[angle=0,width=0.5\textwidth]{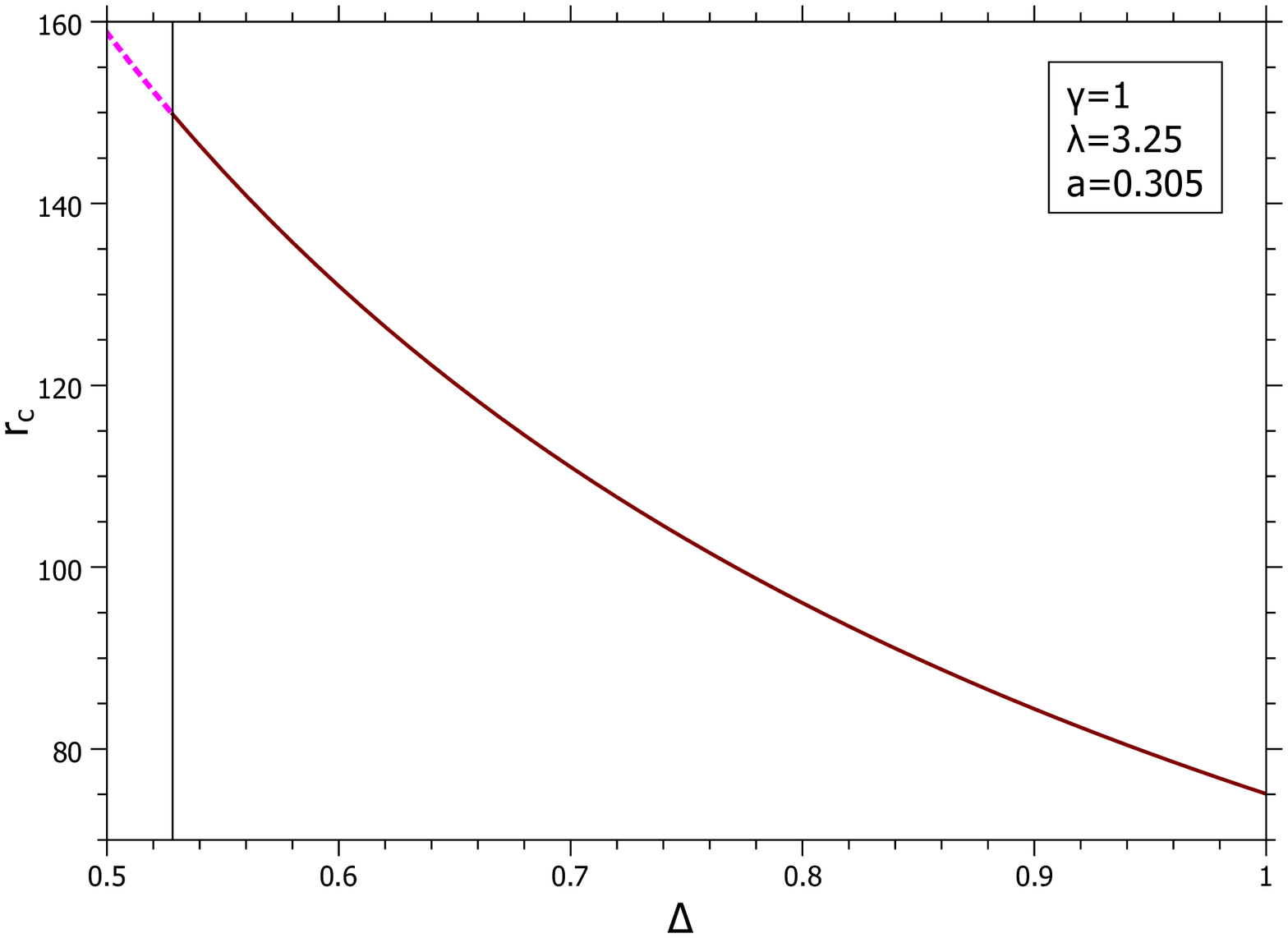}
\caption{For CO isothermal flow, the locus of the position of the single critical point, dotted (Magenta) line (left to the vertical line) and of the outermost critical point, solid (Brown) line (right to the vertical line), for varying $ \Delta $. Here the chosen parameter values are $ \gamma=1 $, $ T \simeq 10^{11}K $, $ \lambda=3.25 $, $ a=0.305 $. The vercical (Black) line separets the Multitransonic region (right) and lone critical point region (left). }\label{f14}
\end{figure}

\begin{figure}[H]
\centering
\includegraphics[angle=0,width=0.5\textwidth]{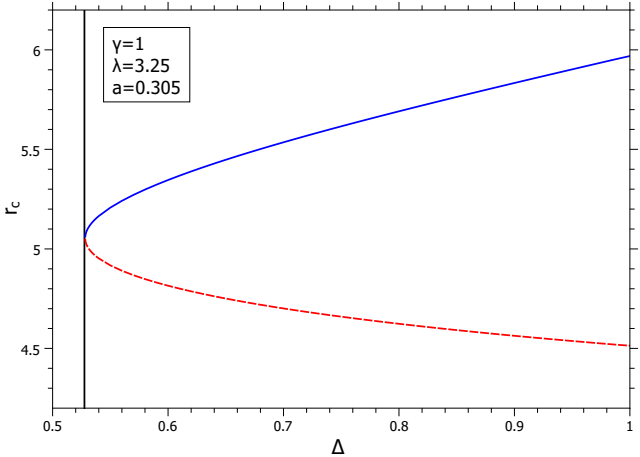}
\caption{For CO isothermal flow, the locus of the position of the middle critical point (the upper arm of the cusp, solid (Blue) line) and innermost critical point (the lower arm of the cusp, dashed (Red) line), for varying $ \Delta $. Here the chosen parameter values are $ \gamma=1 $, $ T \simeq 10^{11}K $, $ \lambda=3.25 $, $ a=0.305 $. The two tracks merges at ($ r_{c} \approx 5.054  $, when $ \Delta=0.5283 $) the vertical line (Black line), and thus it is boundary of the Multitransonic region (right)}\label{f15}
\end{figure}

\section{INTRODUCTION OF TIME DEPENDENT PERTURBATION}
\paragraph*{}
The time-dependent generalization of the continuity condition for an axisymmetric accretion disc in fractal medium, is given by equation \eqref{eq.4}. Substituting the disc surface density expression, $ \Sigma \simeq \rho H $, one can, therefore, obtain,
\begin{equation}\label{eq.401}
\frac{\partial \rho }{\partial t} + \frac{1}{r^{(2\Delta-1)} H} \frac{\partial}{\partial r} (\rho H v r^{(2\Delta-1)}) = 0 .
\end{equation}

As prescribed in literature (Roy \& Ray 2007\cite{Roy et al. 2007}, Bhattacharjee et al. 2009\cite{Bhattacharjee et al. 2009}, Nag et al. 2012\cite{Nag et al. 2012}), here also for the fractal flow system, define a new physical variable $ f = \rho H v r^{(2\Delta-1)} $. In particular, it's stationary value $ f_{0} $ is a constant, which is very much connected to the matter flow rate. For CH and CO models $H$ is either constant(\autoref{eq.14}) or linear function of $r$ only (\autoref{eq.15}). In these systems, a time dependent perturbative analysis is introduced according to the scheme, $ v(r,t) = v_{0}(r) + \tilde{v}(r,t) $, $ \rho(r,t) = \rho_{0}(r) + \tilde{\rho}(r,t) $ and $ f(r,t) = f_{0}(r) + \tilde{f}(r,t) $ (here the subscript 0 denotes stationary background values and tilde implies for perturbed quantities, for all the cases). Therefore, the perturbation is to be seen as a disturbance on the steady, constant background accretion rate.

The definition of $ f $ will lead to a linearised dependence among $ \tilde{f}, \tilde{\rho} $ and $ \tilde{v} $, as, 

\begin{equation}\label{eq.402}
\frac{\tilde{f}}{f_{0}} = \frac{\tilde{\rho}}{\rho_{0}} + \frac{\tilde{v}}{v_{0}} ,
\end{equation}

which is a relation that connects all the three fluctuating quantities, $ \tilde{v} $, $ \tilde{\rho} $ and $ \tilde{f} $, with one another. Going back to equation \eqref{eq.401}, it becomes possible to connect $ \tilde{\rho} $ exclusively to $ \tilde{f} $, through the relation

\begin{equation}\label{eq.403}
\frac{\partial \tilde{\rho} }{\partial t} + \frac{v_{0} \rho_{0}}{f_{0}} \left( \frac{\partial \tilde{f} }{\partial r} \right)  = 0 .
\end{equation}

In the case of VE flow model, $H$ is dependent both on $\rho$ and $r$ according to the relation, eq.~\eqref{eq.16} and hence the equation~\eqref{eq.402} will get modified as,
\begin{equation}\label{eq.404}
\frac{\partial}{\partial t} \left[  \rho^{(\gamma+1)/2}  \right] + \frac{\sqrt{\phi'}}{ r^{(2\Delta-1/2)} } \frac{\partial}{\partial r} \left[  \frac{ \rho^{(\gamma+1)/2} v r^{(2\Delta-1/2)} }{\sqrt{\phi'}} \right]  = 0 .
\end{equation}

from which, it becomes,
 \[ f = { \rho^{(\gamma+1)/2} v r^{(2\Delta-1/2)} } / {\sqrt{\phi'}} \] 
and
\begin{equation}\label{eq.405}
\frac{\tilde{f}}{f_{0}} = \left( \frac{\gamma+1}{2} \right) \frac{\tilde{\rho}}{\rho_{0}} + \frac{\tilde{v}}{v_{0}} ,
\end{equation}

From equation \eqref{eq.404}, it is also very easy to set down the density fluctuations, $ \tilde{\rho} $, in terms of $ \tilde{f} $, as
\begin{equation}\label{eq.406}
\frac{\partial \tilde{\rho} }{\partial t} + \beta^{2} \frac{v_{0} \rho_{0}}{f_{0}} \left( \frac{\partial \tilde{f} }{\partial r} \right)  = 0 .
\end{equation}

with $ \beta^{2} = 2(\gamma+1)^{-1} $, as before. This result may be compared with equation \eqref{eq.403} and the difference is noted. If, however, one were to study an isothermal flow balanced by hydrostatic equilibrium in the vertical direction, then equation \eqref{eq.16} would have to be constrained by $ \gamma=1 $ and $ c_{s} $ being constant. Under these conditions, the expression for density fluctuations in the flow will be identical to equation \eqref{eq.403}, rather than be described by equation \eqref{eq.406}.

\paragraph*{}
The velocity fluctuation comes out to be,
\begin{equation}\label{eq.407}
\frac{\partial \tilde{v} }{\partial t} = \frac{v_{0}}{f_{0}} \left( \frac{\partial \tilde{f} }{\partial t} + v_{0} \frac{\partial \tilde{f} }{\partial r} \right),
\end{equation}

which, upon a further partial differentiation with respect to time, will give

\begin{equation}\label{eq.408}
\frac{\partial^{2} \tilde{v} }{\partial t^{2}} = \frac{\partial}{\partial t} \left[ \frac{v_{0}}{f_{0}} \left( \frac{\partial \tilde{f} }{\partial t} \right)  \right] + \frac{\partial}{\partial t} \left[ \frac{v_{0}^{2}}{f_{0}} \left( \frac{\partial \tilde{f} }{\partial r} \right)  \right] .
\end{equation}

The time-dependent equation for the radial drift is given as equation \eqref{eq.10}, from which the linearised fluctuating part could be extracted as

\begin{equation}\label{eq.409}
\frac{\partial \tilde{v} }{\partial t} + \frac{\partial}{\partial r} \left( v_{0} \tilde{v} + c^{2}_{s0} \frac{\tilde{\rho}}{\rho_{0}} \right) = 0 ,
\end{equation}

with $ c_{s0} $ being the speed of sound in the steady state. Differentiating equation \eqref{eq.409} partially with respect to $ t $, and making use of either equation \eqref{eq.403} or \eqref{eq.406}, along with equation \eqref{eq.407} and \eqref{eq.408}, to substitute for all the first and second-order derivatives of $ \tilde{v} $ and $ \tilde{\rho} $, will deliver the result

\begin{eqnarray}\nonumber\label{eq.410}
\frac{\partial}{\partial t} \left[ \frac{v_{0}}{f_{0}} \left( \frac{\partial \tilde{f} }{\partial t} \right)  \right] + \frac{\partial}{\partial t} \left[ \frac{v_{0}^{2}}{f_{0}} \left( \frac{\partial \tilde{f} }{\partial r} \right)  \right] +\frac{\partial}{\partial r} \left[ \frac{v_{0}^{2}}{f_{0}} \left( \frac{\partial \tilde{f} }{\partial r} \right)  \right] 
\\ 
+ \frac{\partial}{\partial r} \left[ \frac{v_{0}}{f_{0}}(v^{2}_{0} - \varpi c^{2}_{s0}) \frac{\partial \tilde{f} }{\partial r} \right] = 0 .
\end{eqnarray}

in which either $ \varpi=1 $ or $ \varpi=\beta^{2} $, depending on the choice of a particular disc geometry and the equation of state used. For isothermal flows, $ \varpi=1 $, for whatever disc geometry one considers CH, CO or VE. The same value of $ \varpi=1 $ is also obtained for polytropic flows in the first two cases (CH, CO) of the height function, H. The common feature running through all these cases is that H in equation \eqref{eq.401} does not have any dependence on time. It is only when the flow is polytropic and the disc height geometry is expressed by VE model, will H have a time-dependence, whose ultimate consequence will be that $ \varpi=\beta^{2} $ in equation \eqref{eq.410}.

The equation~\eqref{eq.410} can be recast into more compact form (as shown in \cite{Chaudhury et al. 2006, Nag et al. 2012}), \begin{equation}
\partial_{\mu}(\mathrm{f}^{\mu\nu}\partial_{\nu})\tilde{f}=0;\label{tildfeq}
\end{equation}
where $\mu,\nu=0,1$ \[\mathrm{f}^{\mu\nu}\equiv \frac{v_0}{f_0}\left(\begin{array}{cc}
1 & v_0 \\ 
v_0 & v_0^2-\varpi c_{s0}^2
\end{array}  \right).\] From this expression following the procedure depicted in refs.~\cite{Chaudhury et al. 2006, Nag et al. 2012}, one may construct the effective space-time metric as envisaged by the acoustic disturbance on the disc as, \begin{equation}
\mathrm{g}_{\textrm{eff}}^{\mu\nu}=\left(\begin{array}{cc}
1 & v_0 \\ 
v_0 & v_0^2-\varpi c_{s0}^2
\end{array}\right).\label{gmunu}
\end{equation}

\paragraph*{}
By a pertubative time-dependent stability analysis of stationary solutions for axisymmetric, inviscid inflows in continuum (Nag et al. 2012\cite{Nag et al. 2012}), one gets exactly same result as equation \eqref{eq.410}, which is for accretion of fractal matter onto a rotating black hole. A noticeable point is that, the general form of the equation of motion, corresponding to the dynamics of such perturbations and hence the related acoustic metric, are different for flow geometric model. As expected it is independent of the nature of the space-time (i.e. the black hole potential), it is due to the fact that the driving potential itself has no time dependence. Detailed solutions of the equation of motion (equation \eqref{eq.410}), for such time dependent linear perturbation scheme, on inviscid, axisymmetric non-fractal flows, has already given in literature (Ray 2003a\cite{Ray 2003a}, Chaudhury et al. 2006\cite{Chaudhury et al. 2006}) and it can also be extended here (Roy 2007\cite{Roy 2007}).

\section{ACOUSTIC SURFACE GRAVITY}
\paragraph*{}
The acoustic surface gravity $ \kappa $ for the stationary background fluid accreting under the influence of post Newtonian black hole potential can be obtained as (see ref. Bili\`{c} et al. 2014\cite{Bilic et al. 2014})

\begin{eqnarray}\nonumber\label{eq.501}
\kappa = \Biggr| \sqrt{(1+2\phi(r)) \left( 1 - \frac{\lambda^{2}}{r^{2}} - 2\phi(r)\frac{\lambda^{2}}{r^{2}} \right) } 
\\
\left( \frac{1}{(1-c^{2}_{sc})} \left[ \frac{dv}{dr}\Biggr|_{r_{c}} - \frac{dc_{s}}{dr}\Biggr|_{r_{c}} \right]  \right) \Biggl| .
\end{eqnarray}

\paragraph*{}
As it is obvious from the explicit expression of $ \kappa $, the surface gravity is a function of our five parameter initial
boundary condition governing the flow, i.e., $ \kappa\equiv\kappa[\gamma,\mathcal{E},\lambda,a,\Delta] $. Now we want to study the dependence of the acoustic surface gravity on black hole spin for fluid flow in a fractal medium. To do this we have to calculate $ \kappa $ for varying $ a $ (within a range of values) with different values of $ \Delta $, keeping the other parameters $ [\gamma,\mathcal{E},\lambda] $ fixed. This $ \kappa-a $ relationship could be demonstrated for both of the sonic points (inner and outer critical points).  However, $ \kappa $ at the outer acoustic horizon has numerical value way less compared to that of calculated at the inner acoustic horizon. This is a generic property (that $ \kappa_{in}\gg\kappa_{out} $) found independent of the nature of the background space-time metric, geometric configuration of the accretion flow, as well as the thermodynamic equation of state used to describe the matter flow in general. This indicates that the numerical value of the acoustic surface gravity correlates with the strength of the gravitational attraction of the background gravitational field. Also to mention in this context that $ \kappa_{out} $ is not sensitive enough on $ a $ when evaluated at the outer acoustic horizon. This is intuitively obvious because at the outer acoustic horizon (which forms a large distance away from the black hole), space-time becomes asymptotically flat and the effect of the black hole spin does not really affect the dynamics of the flow, and hence the nature of the sonic geometry embedded within. 

\begin{figure}[H]
\centering
\includegraphics[angle=0,width=0.5\textwidth]{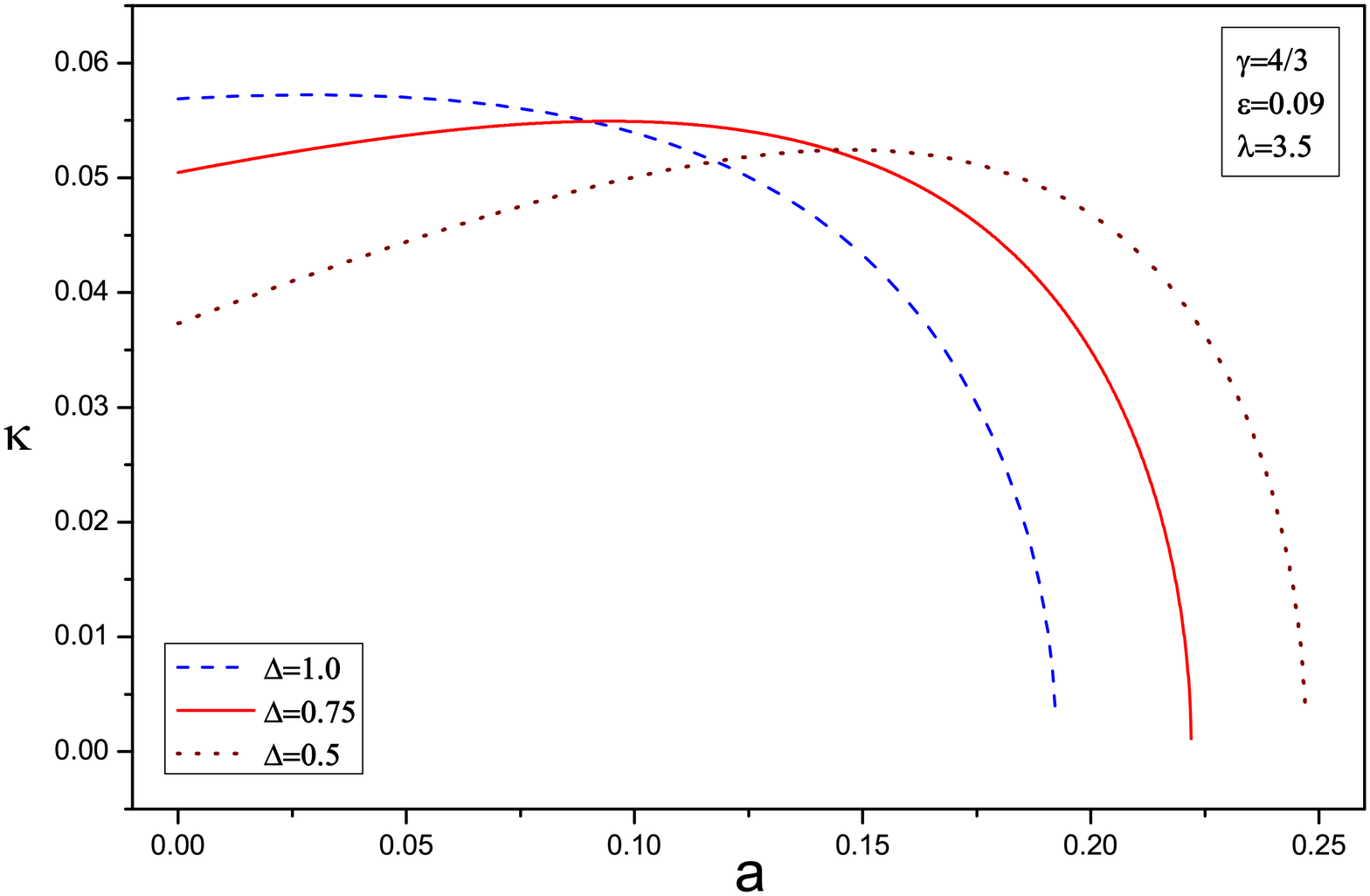}
\caption{For mono-transonic accretion in CO polytropic flows, the $ \kappa-a $ dependence with different values of $ \Delta $, (see text). Here the chosen parameter values are $ \gamma=4/3 $, $ \mathcal{E}=0.09, \lambda=3.5 $.}\label{surfacegravity_a_l3pt5_co}
\end{figure}

\begin{figure}[H]
\centering
\includegraphics[angle=0,width=0.5\textwidth]{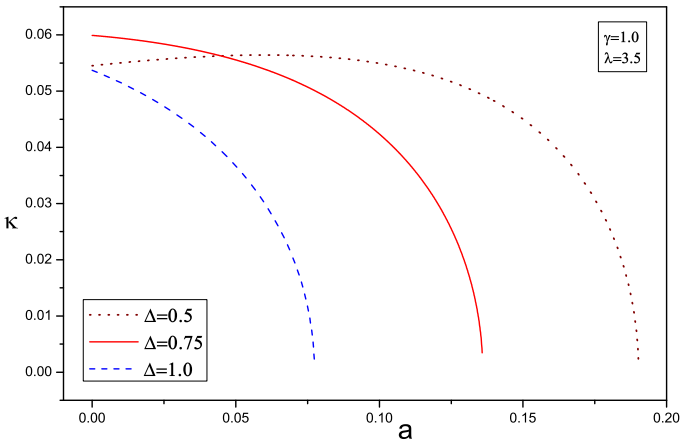}
\caption{For mono-transonic accretion in CO isothermal flows, the $ \kappa-a $ dependence with different values of $ \Delta $ (see text). Here the chosen parameter values are $ \gamma=1.0 $, $ T=9.39 \times 10^{10} K, \lambda=3.5 $.}\label{surfacegravity_a_l3pt5_co_iso}
\end{figure}

\paragraph*{}
Considering the conical geometry, for a fixed subset of parameter space, [$ \mathcal{E}, \lambda $] (polytropic flows) or [$ T, \lambda $] (isothermal flows), where there exists only one critical point, the plots of \autoref{surfacegravity_a_l3pt5_co} (for polytropic flows) or \autoref{surfacegravity_a_l3pt5_co_iso} (for isothermal flows), show, the variation of acoustic surface gravity $ \kappa $, with black hole spin parameter $ a $, for different values of fractal parameter $ \Delta $. In these, $ \kappa-a $ plots, the dashed (Blue) line is for $ \Delta=1.0 $, the solid (Red) line is for $ \Delta=0.75 $ and the dotted (Brown) line is for $ \Delta=0.5 $. This shows, in mono-transonic region, how the $ \kappa-a $ dependence changes, when the fluid medium becomes more and more clumpish (i.e. fractal nature increases), from a continuous medium ($ \Delta=1.0 $).

\begin{figure}[H]
\centering
\includegraphics[angle=0,width=0.5\textwidth]{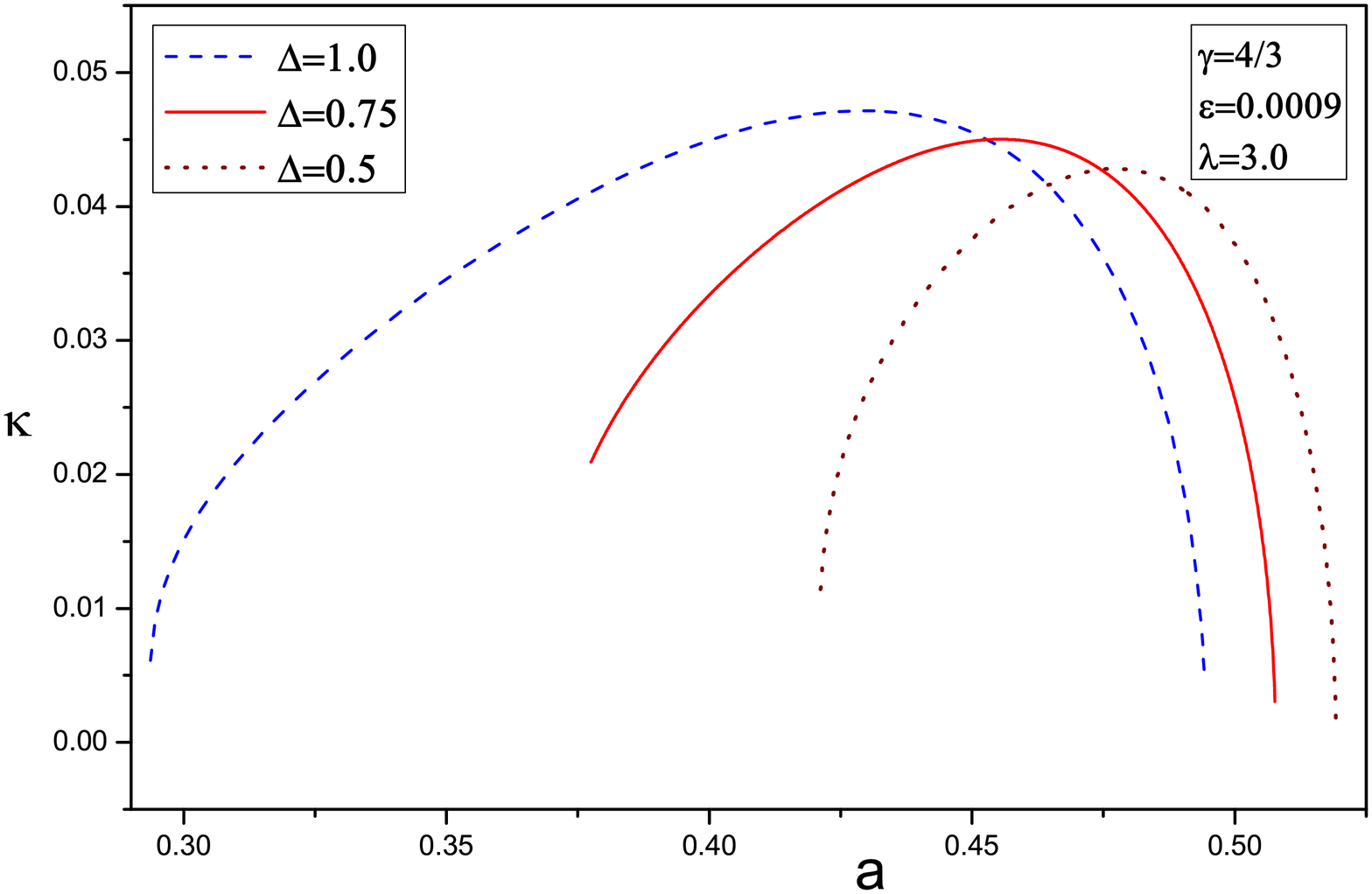}
\caption{For polytropic flows in CO model, the $ \kappa-a $ dependence with different values of $ \Delta $, at inner acoustic horizon (see text). Here the chosen parameter values are $ \gamma=4/3 $, $ \mathcal{E}=0.0009, \lambda=3.0 $.}\label{surfacegravity_a_co}
\end{figure}

\begin{figure}[H]
\centering
\includegraphics[angle=0,width=0.5\textwidth]{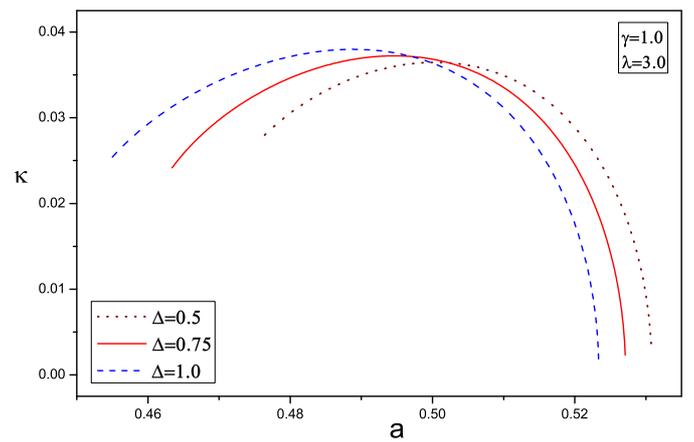}
\caption{For isothermal flows in CO model, the $ \kappa-a $ dependence with different values of $ \Delta $, at inner acoustic horizon (see text). Here the chosen parameter values are $ \gamma=1.0 $, $ T=9.39 \times 10^{10} K, \lambda=3.0 $.}\label{surfacegravity_a_co_iso}
\end{figure}

\paragraph*{}
Considering the conical geometry, \autoref{surfacegravity_a_co} (for polytropic flows) and \autoref{surfacegravity_a_co_iso} (for isothermal flows), show the variation of acoustic surface gravity $ \kappa $ (evaluated at the inner transonic critical point), with black hole spin parameter $ a $, for different values of fractal parameter $ \Delta $. Here the chosen parameter space subset is such that, it produces multitransonic accretion flow. In $ \kappa-a $ plots, the dashed (Blue) line is for $ \Delta=1.0 $, the solid (Red) line is for $ \Delta=0.75 $ and the dotted (Brown) line is for $ \Delta=0.5 $. This shows, for the inner sonic point, how the $ \kappa-a $ dependence changes, when the fluid medium becomes more and more clumpish (i.e. fractal nature increases), from a continuous medium ($ \Delta=1.0 $).

\section{DISCUSSIONS AND CONCLUSIONS}
\paragraph*{}
For a fixed set of values of $ [\gamma,\mathcal{E},\lambda] $, the corresponding numerical domain of [$ a-\Delta $], for which the multi-criticality has been observed, is different for different geometrical configuration associated with accretion disc structures. A fractal medium can be considered as a continuum with the effective density lower than the exact continuum. As a result, the pressure build up against gravity in a fractal medium, is lesser in comparison with continuous medium. In a fractal medium, it is easy to accrate mass through accretion than to make a out flow by wind. The multi-critical numerical domain of [$ \gamma, \mathcal{E}, \lambda, a $], is not only different for different disc geometry models but also different for different fractal parameter.

\paragraph*{}
The variation of $ \Omega^{2} $ with fractal parameter $ \Delta $ keeping other parameters fixed, gives the physical insight of the mathematical nature of critical points affected by the density distribution of the flow medium. This have been shown by \autoref{omega2_del} for polytropic flow or \autoref{omega2_del_iso} for isothermal flow. 

\paragraph*{}
\autoref{f8} (for polytropic case) and \autoref{f14} (for isothermal case) show that the position of the outer critical point (Brown solid line) get shifted outwards monotonically with decreasing Fractal parameter \textquotedblleft{$ \Delta $}\textquotedblright. As maintioned earlier, $ \Delta \longrightarrow 1 $ is the limiting condition of a fractal medium to be a continuous medium, and when $ \Delta $ decreases more, fractal properties become more pronounced. To have accretion feasible, gravity have to overcome the pressure build up of the infilling gas. This pressure depends on the density of the fluid medium through the polytropic equation of state. A fractal medium can be considered as a continuum with the effective density lower than the exact continuum. As a result, the pressure build up against gravity in a fractal medium, is lesser in comparison with continuous medium. In this situation a fractal medium with lower $ \Delta $ has more dilute fluid in comparison with a medium with higher $ \Delta $ and as a consequence can produce lesser pressure. So, in a lower $ \Delta $ medium gravitational pull can accrate mass even from a greater distance. This is the reson, for which the transonic scale-length becomes larger and lager for a more fractal medium (Roy \& Ray 2009\cite{Roy et al. 2009}).

\paragraph*{}
The nonlinear, non-monotonic $ \kappa-a $ dependence, attains a maximum $ \kappa $ value, at particular value of $ a $. In each of these cases, this maxima of $ \kappa-a $ profile, shifts towards the higher $ a $ values, as the fractal property of the flow medium becomes more stronger. From various surface gravity plots, it can be concluded that, the overall variation of $ \kappa-a $ profiles, for the change in the fractal nature of the flow medium, are quantitatively similar at both mono-transonic and inner transonic (at inner acoustic horizon in multitransonic regime) critical point. Thus, the study of accretion of fractal matter onto a rotating black hole and evaluation of emergent acoustic surface gravity has been successfully investigated in this work.

\section*{ACKNOWLEDGMENTS}
Several visits of SN at HRI has been supported by the Cosmology and High Energy Astrophysics project funding. Visit of SM at HRI was supported under the Visiting Student Programme (VSP) scheme of HRI. The present work is an extension of the M.Sc. thesis completed by SM under the supervision of SN utilizing the facilities provided by the Department of Physics, Sarojini Naidu College for Women, Kolkata.

\end{multicols}

\end{document}